\documentclass[12pt, english]{article}
\usepackage{amsmath,amssymb,color,float}
\usepackage{stmaryrd}
\usepackage{graphicx,psfrag,epsf}
\usepackage[authoryear]{natbib}
\usepackage{fullpage,setspace}
\usepackage{subcaption}
\usepackage{changepage}
\usepackage{listings}
\usepackage{colortbl}
\usepackage{xcolor}
\usepackage{ragged2e}
\usepackage{tabularx}
\usepackage{authblk}
\usepackage{booktabs}

\usepackage{amssymb}

\usepackage{lineno}

\title{Accounting for variable detection functions in temporal abundance modeling via transfer learning}

\author[1]{Kevin M. Collins}
\author[1]{Erin M. Schliep}
\author[2]{Tyler Wagner}
\author[3]{Christopher K. Wikle}
\affil[1]{Department of Statistics, North Carolina State University, North Carolina, U.S.A.}
\affil[2]{Department of Biology, Miami University, Oxford, OH, U.S.A.}
\affil[3]{Department of Statistics, University of Missouri, Columbia, Missouri, U.S.A.}

\begin{document}


\maketitle

\begin{abstract}
Relative abundance, measured as the number of animals caught per unit of sampling effort (CPUE), is commonly used to monitor fish and wildlife populations, largely because sampling methods are cost-effective to implement. Modeling relative abundance, however, requires the assumption that the detection probability is constant across sampling events. This assumption is likely not valid, as the probability of detection often varies as a function of several factors, including the characteristics of individual animals and environmental conditions at the time of sampling. In contrast, methods to estimate absolute abundance, such as capture-recapture (CR), account for variable detection, but are often infeasible to implement across large spatiotemporal scales. Despite this, CR data are sometimes available for species of interest, albeit at smaller spatiotemporal extents. Leveraging information on detection probabilities from CR data to help inform estimates of widely available CPUE data could strengthen inferences about the status of fish and wildlife populations. We propose an approach to (i) learn the effect of environmental covariates on detection probabilities from CR data and (ii) transfer these detection functions to CPUE models for improved inference. Shown empirically through a simulation study, this approach improves estimates of abundance and the ability to detect temporal trends. We apply our transfer learning method using CR and CPUE data to recreationally important smallmouth bass (\textit{Micropterus dolomieu}) fisheries in Pennsylvania, USA rivers.
\end{abstract}


\textbf{Open Research statement:} The code for model fitting is available at \\ https://doi.org/10.5066/P1XJ5UDV. The data underlying this article were provided by the Pennsylvania Fish \& Boat Commission by permission. Data can be obtained by request to G.D. Smith geofsmith@pa.gov. \\

\section{Introduction}

The management of fish and wildlife populations relies on knowledge of their abundance to assess status and trends. Although specific monitoring designs for obtaining information about population size vary as a function of the target species, management and conservation objectives, and available resources \citep{allen2015evaluating}, they typically obtain information about population size using either an index of abundance or by estimating absolute (true) abundance. Indices of abundance typically consist of the number of animals caught per unit of sampling effort (catch per unit effort; CPUE), where effort can be recorded differently depending on the species, habitat, and methods used (e.g., hours sampled, or volume or area of habitat surveyed). Indices of abundance are often used to monitor fish and wildlife populations and this reliance on using indices of abundance to make inferences on the status and trends of ecologically and socioeconomically important species is largely because sampling methods are cost-effective to implement \citep{ye2009reliable, ducharme2018indices}. This is contrast to methods that estimate true abundance, which are more expensive and time consuming to implement -- often requiring multiple visits to a single site over time \citep[e.g., capture-recapture (CR) methods; ][]{seber1965note, jolly1982mark} making them infeasible to implement across large spatiotemporal extents. Importantly,  CPUE methods assume that the index of abundance is proportional to the true abundance and that the probability of successfully capturing/counting any one individual is perfect or constant over space and time. Because the probability of capturing or counting (termed catchability or detection probability for capture and count sampling methods, respectively) an individual is most often less than one and varies over space and time as a function of environmental conditions \citep{ kellner2014accounting, north2024accounting}, indices like CPUE are an imperfect measure of true abundance. Failing to account for imperfect detection can lead to biased estimates and incorrect inferences about population status and trends. 

One approach monitoring programs use to try to minimize the influence of imperfect and varying detection is to implement a design that attempts to keep detection probability relatively constant over space and time \citep{pollock2002large}. This may involve sampling populations at the same time and under similar environmental conditions each year. However, attempts to standardize the timing and environmental conditions of sampling may not be adequate to control for imperfect and varying detection and CPUE data may need to be corrected for imperfect detection before being used as reliable indicators of abundance \citep{staton2022accounting}. Approaches to correct CPUE data are particularly important for inland freshwater fisheries that are critically important for global food security, local and regional economies, and provide important recreational and commercial fisheries \citep{funge2019fresh, murray2020trends}. However, monitoring the vast number of inland freshwater fisheries is challenging and makes CR methods difficult to operationalize at a management relevant scale.

As such, we explore the possibility of leveraging information in CR data to improve inference based on CPUE data, which falls under the notion of transfer learning \citep{weiss2016survey}. Transfer learning describes a setting in which we wish to learn about one population (the target population) by leveraging information from a related, but perhaps not identical, population (the source population) to improve inference. Typically, transfer learning is employed when the source population is significantly larger than the target population. In our case, it is the opposite, where the data on the source population are scarce, yet more rich in information, meaning the benefits of transfer still apply. Bayesian modeling is a very natural framework in which to transfer information, a review of which can be found in \cite{suder2023bayesian}.

The idea of combining data that informs detection probabilities and count data is not new, and has been explored in the context of ``sightability" models \citep{fieberg2012estimating}. In these types of models a ``sightability" trial is conducted on the population of interest to estimate detection probabilities, and then these probabilities are inserted into a Horvitz-Thompson estimator to get estimates of abundance. \cite{ver2025integrated} extended this approach to a hierarchical Bayesian context for an analysis of harbor seal populations. Their approach is similar to ours, but in the spirit of transfer learning, we consider scenarios in which the assumption of population homogeneity is violated, i.e. the source and target populations have different detection functions. This explores an important notion of negative transfer \citep{wang2019characterizing}, wherein the violation of a transferability assumption can possibly result in ``worse" inference on the target population. Furthermore, we adhere to the traditional CPUE approach of modeling counts as Poisson-distributed with detection probability being a modifier similar to effort, rather than directly modeling a Binomial count as in the sightability model. \citet{hooten2007hierarchical} similarly sought to account for the effect of detection probability on abundance modeling across heterogeneous populations by assigning a prior distribution to the detection probability of Eurasian Collared Doves that was obtained from a repeated observation study of a related species, the Mourning Dove.  This type of prior elicitation is a form of transfer learning, but their analysis assumed constant detection probability, with no covariate effects, which is a potentially serious limiting assumption that we seek to overcome. Opening up the possibility of transferring information across heterogeneous populations (i.e. different species) could improve monitoring capabilities for populations that are typically only monitored via count data due to financial and labor constraints.

The remainder of this paper is structured as follows. In Section 2, we establish the necessary background on how CR and CPUE data are typically modeled, which then motivates our proposed transfer learning approach. We demonstrate the efficacy of our approach in Section 3 under a number of different settings that reflect the noisiness of the detection function and the appropriateness of the transferability assumption. Finally, in Section 4 we apply our method to monitoring data collected from economically and recreationally important smallmouth bass (\textit{Micropterus dolomieui}) fisheries across Pennsylvania, USA rivers, and then conclude with a discussion.

\section{Methodology}

\subsection{Population Model}\label{pop_model}

For a multivariate population over time, let $N_{tj}$ denote the true population of type $j$ (e.g., species, size class) for year $t$. We assume this population is fixed for year $t$ while sampling may occur at multiple time points within the year. In our analysis, we are interested in $J$ different size classes for the same species so henceforth ``type" is referred to as size class, but this could be any labeling structure of interest. For us, the total population across all size classes for year $t$ can be denoted $N_t=\sum_{j=1}^JN_{tj}$. Let $\textbf{N}$ denote the vector consisting of all $N_{tj}$, for all observed times $t$ and $j=1, \dots J$.

The true population of each size class is modeled as
\begin{equation}
    N_{tj}\sim Pois(\theta_{tj}),
\end{equation}
where $\theta_{tj}$ denotes the expected abundance of size class $j$ in year $t$. We model the vector $\boldsymbol{\theta}_t = (\theta_{t1}, \dots, \theta_{tJ})'$ jointly on the log scale as
\begin{equation}\label{theta_model}
    \begin{pmatrix}
    \log\theta_{t1} \\ \vdots \\ \log\theta_{tJ}
\end{pmatrix}
\sim  N (
\textbf{A}\textbf{z}_t\boldsymbol,
\boldsymbol{\Omega}
)
\end{equation}
Here, $\textbf{z}_t$ is a $q_z$-dimensional vector of year-specific covariates and $\textbf{A}$ is a $J\times q_z$ coefficient matrix where the $j$th row, denoted $\boldsymbol\alpha_j'$, is a vector of coefficients for size class $j=1,\dots,J$. Additionally, $\boldsymbol\Omega$ is a $J \times J$ covariance matrix that allows for stochastic dependence between the $J$ size classes. Note that if $j$ denotes species, this is analogous to a joint species distribution model specification \citep{ovaskainen2020joint}.

Given this hierarchical model framework, inference on the model parameters requires data, i.e., observations of $N_{tj}$. Barring a complete census, $N_{tj}$ cannot be observed directly and instead must be inferred via an incomplete sample. Below we outline two different observation types, namely, capture-recapture data and catch per unit effort data, and the resulting modeling approaches for linking the observations to the processes and parameters of interest. 

\subsection{Capture-recapture}\label{cr_model}

Capture-recapture data are obtained from repeat sampling of a population across multiple sampling occasions. Here, we assume sampling occurs on multiple days (possibly non-consecutive) within each year. Let $d_t$ denote the number of days sampled in year $t$. We let $n_{tjk}$ denote the total number of fish caught in year $t$ of size class $j$ on day $k$ and $m_{tjk}\leq n_{tjk}$ denote the number of those fish that were previously marked. For each year, we assume that $m_{tj1}=0$, meaning no fish were marked prior to the first sampling day such that $n_{tj1}$ consists only of unmarked fish. On the $k^{th}$ day, $n_{tjk}$ fish of size class $j$ are caught and subsequently released after the $n_{tjk}-m_{tjk}$ unmarked fish are all marked. Thus, our data consist of the tuples $\{n_{tjk}-m_{tjk},m_{tjk}\}$ for $j=1,\dots,J$, $t=1,\dots,T$, and $k=1,\dots,d_t$. Let $M_{tjk}=\sum_{l=1}^{k-1} n_{tjk}-m_{tjk}$ for $k>1$ and $M_{tj1}=0$ denote the total marked population available to be caught on a given day $k$.

We assume that $\{n_{tjk}-m_{tjk},m_{tjk}\}$ are sampled from their respective populations $\{N_{tj}-M_{tjk},M_{tjk}\}$ with equal probability $p_{tjk}$. That is, unmarked and marked fish are (re)captured with equal probability. We refer to $p_{tjk}$ as the detection probability for year $t$ of size class $j$, on day $k$. Let $(\textbf{n-m},\textbf{m})$ denote the matrix of data across all years and size classes and $\textbf{p}_{tj}$ the vector of all probabilities for year $t$ and size class $j$ respectively. Following \cite{castledine1981bayesian}, we assume the likelihood
\begin{equation}
L(N_{tj},\textbf{p}_{tj};(\textbf{n-m},\textbf{m}))\propto {N_{tj} \choose \sum_k n_{tjk} - \sum_k m_{tjk}}\prod_{k=1}^{d_t} p_{tjk}^{n_{tjk}}(1-p_{tjk})^{N_{tj}-n_{tjk}}.
\end{equation}
Note that $N_{tj}$ is the unobserved true abundance and is thus treated as a latent parameter to be inferred given the data.

We model the size class detection probabilities for a given year and day as functions of covariates on the day of sampling. Let  $\textbf{x}_{tk}$ denote a $q_x$-dimensional vector of covariates that are assumed to capture the variation in detection across years and day of year. Using a logit link function, we model
\begin{equation}
\text{logit}(p_{tjk}) = \textbf{x}_{tk}'\boldsymbol\beta_j + \epsilon_{tjk},
\end{equation}
where $\epsilon_{tjk}\sim N(0,\sigma^2_j)$ are independent random effects. The regression coefficients, $\boldsymbol\beta_j$, and variance parameters, $\sigma^2_j$, are size class specific.

In a Bayesian framework, we assign prior probabilities to each unknown parameter, denoted $\pi(\cdot)$. Thus, the full hierarchical model specification that combines the model for abundance in (\ref{pop_model}) with the likelihood for the CR data results in the posterior distribution
\begin{equation}
    \pi(\boldsymbol\beta,\sigma^2,\textbf{N},\textbf{A},\boldsymbol\Omega|(\textbf{n-m},\textbf{m}))\propto\prod_{j,t}L(N_{tj},\textbf{p}_{tj};(\textbf{n-m},\textbf{m}))\pi(\textbf{p}_{tj}|\boldsymbol\beta,\sigma^2)\pi(\textbf{N}|\textbf{A},\boldsymbol\Omega)\pi(\boldsymbol{\beta},\sigma^2,\boldsymbol{\alpha},\boldsymbol\Omega).
\end{equation}

\subsection{Catch per unit effort}\label{cpue_model}

Here, let $y_{tjk}$ denote the number of fish caught in year $t$, of size class $j$, on day $k$. Importantly, when modeling CPUE data, a measure of effort also needs to be specified.
In contrast to capture-recapture data, catch per unit effort data provide no information with respect to the detection probabilities. The inclusion of effort is meant to standardize the catch magnitudes under different conditions in order to compare relative abundance over time. Typically, effort is assumed to vary as a function of known sampling time (e.g., hours), number of gears deployed (e.g., nets), or personnel \citep{hubert2007relative}, although see \cite{north2024accounting}, meaning it is day, year, and size-class specific.  

CPUE data are typically modeled as
\[
y_{tjk}\sim Pois(\lambda_{tj}e_{tjk})
\]
where $\lambda_{tj}$ defines the latent measure of relative abundance in year $t$ of size class $j$ and $e_{tjk}$ provides an offset term for effort. In our analysis, effort is assumed the same across size class meaning $e_{tjk}=e_{tk}$.

In relating the model in Section \ref{pop_model} to the CPUE model, $\lambda_{tj}$ and $\theta_{tj}$ are both measures of abundance where, in theory, $\lambda_{tj} \propto \theta_{tj}$. When specifying a model for $\lambda_{tj}$, we therefore might assume a similar model to (\ref{theta_model}). Here, we define the joint model for $\lambda_{tj}$ for $j=1, \dots, J$ on the log scale as
\begin{equation}
    \begin{pmatrix}
    \log\lambda_{t1} \\ \vdots \\ \log\lambda_{tj}
\end{pmatrix}
\sim  N (
\textbf{G}\textbf{z}_t,
\boldsymbol{\Sigma}
)
\end{equation}
where $\textbf{G}$ is again a coefficient matrix with the $j$th row, denoted $\boldsymbol{\gamma}'_j$, constructed similar to $\textbf{A}$ above, $\mathbf{z}_t$ is the vector of covariates, and $\boldsymbol{\Sigma}$ is the covariance matrix capturing possible dependence across size classes.
Note that in the ideal situation where $\lambda_{tj}$ and $\theta_{tj}$ differ only in a proportionality constant, the parameters $\boldsymbol\gamma_j$ and $\boldsymbol\Sigma$ have the same interpretation in terms of sign as $\boldsymbol\alpha_j$ and $\boldsymbol\Omega$, but differ in magnitude. 

Then, letting $\textbf{y}$ and $\boldsymbol\lambda$ be the vectors of all $y_{jtk}$ and $\lambda_{tj}$ respectively, we conduct inference via the resulting posterior
\begin{equation}\pi(\boldsymbol\lambda,\textbf{G},\boldsymbol\Sigma|\textbf{y})\propto \prod_{j,t,k}L(\lambda_{jt};y_{tjk})\pi(\boldsymbol\lambda|\textbf{G},\boldsymbol\Sigma)\pi(\textbf{G})\pi(\boldsymbol\Sigma).
\end{equation}

The primary difference in the two modeling approaches is that the CR model allows for inference on the detection function, and therefore absolute abundance. Conversely, the CPUE data provide no insight into detection probabilities and thus no insight into absolute abundance. The CPUE model is not agnostic to detection probability, however, as it implicitly assumes that $y_{tjk}$ fish are sampled from the population of size $N_{tj}$ at a constant rate across sampling occasions, controlling for effort. In fact, it may be more appropriate to describe the assumed data generating mechanism as
\begin{equation}
    y_{tjk}\sim Binomial(N_{tj},p).
\end{equation}
Unfortunately, the absence of recaptures renders this model unidentifiable -- we are unable to learn both $N_{tj}$ and $p$. Instead, we tend to model $y_{tjk}\sim Pois(\lambda_{tj}e_{tjk})$, with known effor $e_{tjk}$. This is satisfactory for drawing inferences on ``relative" abundance as the Poisson is the limiting distribution of a Binomial and therefore, for large $N_{tj}$, a Poisson with rate parameter $\lambda_{tj} \approx N_{tj}p$ is identifiable. Thus, this model specification comes with the strong assumption of constant detection probability, an assumption that is readily invalidated in lotic systems \citep{gwinn2016imperfect} as is the case in our data analysis. This motivates the need for methods that enable inference on changes in abundance based on CPUE data, while abandoning the assumption of constant detection probability, which is the focus of our proposed work.

\subsection{Transfer learning from CR model to CPUE model}\label{adjusted_model}

Given access to CR and CPUE data, we propose a transfer learning approach that improves inference under the CPUE model. This approach allows us to relax the assumption of constant detection probability and instead assume that the detection probability is a function of environmental covariates plus noise. 

Consider two data sources: $(\textbf{n}-\textbf{m},\textbf{m})= \{(n_{tjk}-m_{tjk},m_{tjk}): j=1,\dots,J$, $t\in T_1\subseteq\{1,\dots,T\}, k=1,\dots,d_{t1}\}$, and $\textbf{y}=\{y_{tjk} : j=1,\dots,J$, $t\in T_2\subseteq\{1,\dots,T\}, k=1,\dots,d_{t2}\}$. Note that the years and days within years of sampling need not be the same for the CR and CPUE data. Additionally, let $\boldsymbol\beta=(\boldsymbol\beta_1',\dots,\boldsymbol\beta_J')'$ and $\boldsymbol\lambda=\{\lambda_{tj}\}_{j=1,\dots,J,t\in T_2}$.

Starting with the CR data, we fit the model outlined in Section \ref{cr_model} and obtain the marginal posterior $\pi(\boldsymbol\beta|(\textbf{n}-\textbf{m},\textbf{m}))$. Next, to link the two models, we first reparameterize the data model in Section \ref{cpue_model} for the CPUE data as
\begin{equation}
    y_{tjk}|\lambda_{tj}^*,p_{tjk}\sim Pois(\lambda_{tj}^*e_{tk}p_{tjk}),
\end{equation}
where $\text{logit}(p_{tjk})=\textbf{x}_{tk}'\boldsymbol\beta_j$. A traditional Bayesian approach would have us assign a prior distribution to $\boldsymbol\beta_j$ and derive a posterior. As previously discussed, $y_{tjk}$ alone provides no information with respect to $p_{tjk}$, meaning we cannot learn about $\boldsymbol\beta_j$. The posterior distribution that we can derive, $\pi(\boldsymbol\lambda^*,\boldsymbol\gamma,\boldsymbol\Sigma|\textbf{y})$, can be written as
\begin{equation}
    \pi(\boldsymbol\lambda^*,\boldsymbol\gamma,\boldsymbol\Sigma|\textbf{y})\propto \int \prod_{j,t,k}L(\lambda_{tj}^*,\boldsymbol\beta_{j};y_{tjk})\pi(\lambda_{tj}^*|\boldsymbol\gamma,\boldsymbol\Sigma)\pi(\boldsymbol\gamma)\pi(\boldsymbol\Sigma)\pi(\boldsymbol\beta)d\boldsymbol\beta.
\end{equation}

Using Markov chain Monte Carlo (MCMC) we can obtain samples from this posterior distribution. Specifically, at each iteration we draw a sample from the posterior distribution of $\boldsymbol\beta\sim\pi(\boldsymbol\beta|(\textbf{n-m},\textbf{m}))$ obtained via the CR model. Then,  conditioning on this value, we sample from each of the full conditional distributions for the remaining model parameters. This step results in the transfer learning of the detection probabilities from the CR model to the CPUE model. Importantly, given that posterior draws are taken of $\boldsymbol{\beta}$ at each iteration, we properly propagate the uncertainty in these detection probabilities through to our CPUE model parameters. This approach is similar to that taken in \cite{ver2025integrated}.

\section{Simulation Study}

We conducted simulation studies under several different scenarios to empirically investigate the efficacy of our transfer learning approach. Under each setting, the true population $N_{tj}$ for each year and size class is simulated according to Section \ref{pop_model} and then CR data $(\textbf{n-m},\textbf{m})$ is simulated according to Section \ref{cr_model}. We obtain a CPUE dataset by retaining just the first day of each year of sampling, setting $y_{tj1}=n_{tj1}$, assuming effort to be constant across all sampling occasions, i.e. $e_{tk}=1$ for all $t$ and $k$. We consider only two size classes $j=1,2$.

Further, we specify covariates $\textbf{x}_{tk}$ and $\textbf{z}_t$ as well as values of all model parameters. Specifically, we define $\textbf{x}_{tk}$ to include an intercept term, and two separate day-specific continuous variables. In $\textbf{z}_{t}$, we include an intercept, a linear year effect, and a continuous variable.
We assume the same sampling schedule as the data analyzed in Section \ref{real_data_analysis} with $t\in\{1,\dots,17\}$ and $k=1,\dots,d_t$. For the model parameters, we set $\boldsymbol\alpha_1=(8,0,-2)'$,$\boldsymbol\alpha_2=(6.5,0.05,-1)'$ with the notable difference of size class $j=1$ having no trend in time, while size class $j=2$ has a positive trend. We set $\boldsymbol\Omega_{12}=\boldsymbol\Omega_{21}=0.1$ to allow for mild dependence between size classes and we let $\boldsymbol\Omega_{11}=\boldsymbol\Omega_{22}=1$. More importantly for our purposes, we set $\boldsymbol\beta_1=(-3.5,-2,0.5)',\boldsymbol\beta_2=(-3.5,0,0)'$, which means that detection of size class 1 is a function of environmental covariates, whereas size class 2 is not, which will allow us to discern the effects of our adjustment when detection is purely stochastic and cannot be accounted for by covariates. We consider five simulation scenarios dictated by the variability in detection, i.e. the percent of variation in detection that is explained by $\textbf{x}_{tk}$: (I) $\sigma^2_1=\sigma^2_2=0.1$, (II) $\sigma^2_1=\sigma^2_2=0.5$, (III) $\sigma^2_1=\sigma^2_2=1$, (IV) $\sigma^2_1=0.2,\sigma^2_2=0.8$ , (V) $\sigma^2_1=0.8,\sigma^2_2=0.2$.

To complete the hierarchical model specification, we assign prior distributions to all parameters. We model $\boldsymbol\alpha_j$, 
$\boldsymbol\gamma_j$, and $\boldsymbol\beta_j$ each with independent uninformative normal priors. The covariance parameters $\boldsymbol{\Omega}$ and $\boldsymbol\Sigma$ are assigned independent inverse-Wishart priors with scale matrix $\textbf{I}_J$ and $J+1$ degrees of freedom, where $\textbf{I}_J$ is the $J\times J$ identity matrix. And the size-class specific variance terms, $\sigma^2_j$, are each given non-informative IG$(0.1,0.1)$ priors.

\subsection{Metrics for comparison}

\begin{figure}
    \centering
    \includegraphics[width=1\linewidth]{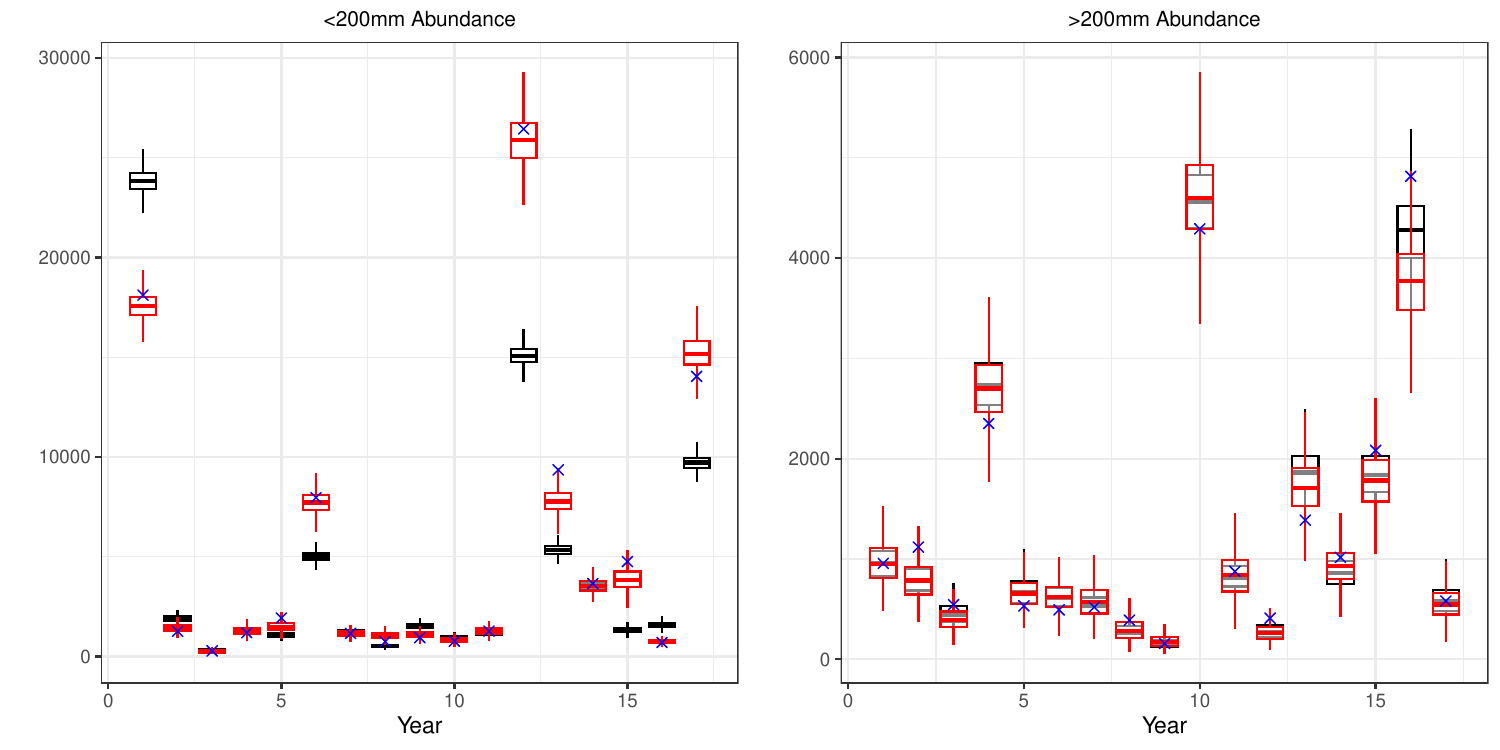}
    \caption{Posteriors for naive ($\tilde N_{tj}$; black) and transfer learning ($\tilde N_{tj}^*$; red) abundance estimates over time for a single simulation run. Estimates are provided for two size classes of fish (size class 1 = $<$ 200mm and size class 2 = $>$ 200 mm total length). The true values of $N_{tj}$ are marked with blue crosses. Note that detection probability is affected by environmental covariates for size class 1, but not for size class 2.}
    \label{fig:sim_example}
\end{figure}

Our modeling goal is to draw inference on changes in the population of interest over time. We define this inferential goal as the ability of our model to recover the true $N_{tj}$ for a given time $t$ and to recover the true trend in $N_{tj}$ over all $t\in\{1,\dots,17\}$. The two model outputs that we aim to compare against true abundance, $N_{tj}$, include our estimates of $\lambda_{tj}$ and $\lambda_{tj}^*$, which we refer to as the naive approach and the transfer learning approach, respectively. The posterior distributions $\pi(\boldsymbol\lambda|\textbf{y})$ and $\pi(\boldsymbol\lambda^*|\textbf{y})$ are not directly comparable given that they are on different scales when $p_{tjk}<1$. To rectify this disparity, we take the implicit assumption of CPUE that $\lambda_{tj}\approx N_{tj}p$ for a constant detection $p$ and transform the posterior samples by dividing $\lambda_{tj}/\hat p_{j}$ where $\hat p_j=\frac{1}{T}\sum_{t}\frac{1}{d_t}\sum_kE[p_{tjk}|(\textbf{n-m},\textbf{m})]$. When comparing the estimates of each model to the true $N_{tj}$, we define new notation for improved clarity. Specifically, define
\begin{equation}
    \tilde N_{tj}=\lambda_{tj}/\hat p_{j}
\end{equation}
\begin{equation}
    \tilde N_{tj}^* = \lambda_{tj}^*
\end{equation}
noting that our model provides a posterior distribution of both $\tilde N_{tj}$ and $\tilde N_{tj}^*$. Figure \ref{fig:sim_example} presents the resulting posterior distributions where the estimates of each of the models are on approximately the same scale facilitating comparison.

To assess the broader trend, we calculate the Mann-Kendall statistic \citep{kendall1948rank} of the true data $N_{tj}$ as well as of the posteriors of $\lambda_{tj}$ and $\lambda_{tj}^*$.
We compute the statistic for each posterior sample, which results in a posterior distribution of Mann-Kendall statistics for each model. The Mann-Kendall statistic is valuable in this setting because it is robust to outliers and nonlinear trends being that it is based on all pairwise differences, and is robust to scale as it is standardized to follow a normal distribution under the null hypothesis being there is no trend. Let $U(N_{tj})$, $U(\tilde N_{tj})$, and $U(\tilde N_{tj}^*)$ denote the statistics for the true population, the naive approach, and the transfer learning approach.

\subsection{Results}
For each of the 5 scenarios we ran 100 simulations and compared the posterior performance of $\tilde N_{tj}$ and $\tilde N_{tj}^*$ in terms of median absolute deviation of the posterior mean and the empirical coverage of the 95\% credible interval of the true $N_{tj}$. Additionally, we consider the empirical coverage of the 95\% credible interval of $U(N_{tj})$ using composition sampling from the posterior.

Table \ref{tab:abundance_performance} shows the average MAD across simulations for each scenario. The MAD for $\tilde N_{t1}^*$ is significantly lower than the MAD for $\tilde N_{t1}$ under all scenarios, illustrating that adjusting for daily variability in detection improves point estimation of abundance. We find that as noise increases, the benefits of the adjustment decreases. Yet, even under scenario V with the largest uncertainty, we see that adjusting under an even noisier detection function still yields more accurate estimates of true population.  Conversely, for size class 2 that had no variability in detection that was attributable to known covariates, the MAD of $\tilde N_{t2}^*$ is approximately equal to the MAD of $\tilde N_{t2}$. 

The 95\% empirical coverages are also reported in Table \ref{tab:abundance_performance}. While the nominal coverage is never attained in any scenario, this can be attributed to the fact that we are not fitting a likelihood that reflects the true data generating distribution, but rather a Poisson that approximates the true likelihood. Empirical coverage is, however, better under the adjustment for both size classes across all simulation settings. Thus, even when there is no known covariate effect on detection, the added uncertainty from the posterior distribution $\pi(\boldsymbol\beta|(\textbf{n-m},\textbf{m)}$ helps account for the inherent variability in the detection probabilities.  

Finally, empirical coverage of $U(N_{tj})$ is reported in Table \ref{tab:mk_cov}. We detect similar improvements for both size classes in each scenario, although the improvement is slightly less dramatic for size class 2. When the detection probability is directly impacted by known covariates, our method demonstrably improves point estimation and coverage of the true abundance, as well as coverage of the true trend in abundance. Benefits still exist for size class 2 since propagating uncertainty in detection, albeit just noise, captures the inherent variability in the detection probabilities that is not accounted for in the naive CPUE model.

\begin{table}[]
\caption{Mean MAD of posterior mean from and 95\% credible interval coverage of true population $N_{tj}$ across simulations under each variance scenario I -- V. $\tilde N_{t1}, \tilde N_{t2}$ denote model output from the naive approach, and $\tilde N_{t1}^*, \tilde N_{t2}^*$ denote model output from the transfer learning approach for size class 1 and 2, respectively. Bold values indicate the best result between naive and transfer learning approaches for a given size class.}
    \centering
    \begin{tabular}{lllcccccccc}
\toprule
& & & 
\multicolumn{4}{c}{MAD from $N_{tj}$} &
\multicolumn{4}{c}{Coverage of $N_{tj}$}  \\
\cmidrule(lr){4-7} \cmidrule(lr){8-11} 
 Scenario & $\sigma^2_1$ & $\sigma^2_2$ & $\tilde N_{t1}$ & $\tilde N_{t1}^*$ & $\tilde N_{t1}$ & $\tilde N_{t1}^*$ & $\tilde N_{t1}$ & $\tilde N_{t1}^*$ & $\tilde N_{t1}$ & $\tilde N_{t1}^*$ \\
 \midrule
I & 0.1 & 0.1 & 1015.16 & \textbf{147.6}5 & \textbf{221.81} & 148.85 & 0.20 & 0.86 & \textbf{0.85} & \textbf{0.93}\\
II & 0.5 & 0.5 & 1268.55 & 353.56 & \textbf{847.45} & \textbf{350.05} & 0.16 & 0.51 & \textbf{0.64} & \textbf{0.75}\\
III & 1.0 & 1.0 & 1791.39 & 601.54 & \textbf{1598.93} & \textbf{592.61} & 0.12 & 0.29 & \textbf{0.65} & \textbf{0.73}\\
IV & 0.2 & 0.8 & 1024.95 & \textbf{511.50} & \textbf{376.06} & 514.36 & 0.19 & 0.36 & \textbf{0.73} & \textbf{0.73}\\
V & 0.8 & 0.2 & 1578.19 & \textbf{188.74} & \textbf{1292.60} & 190.89 & 0.12 & 0.77 & \textbf{0.66} & \textbf{0.88}\\
\bottomrule
\end{tabular}

    \label{tab:abundance_performance}
\end{table}

\begin{table}[]
    \caption{Coverage of 95\% credible interval of Mann-Kendall statistic $U(N_{tj})$ across simulations under each variance setting. $U(\tilde N_{t1}), U(\tilde N_{t2})$ denote Mann-Kendall statistics calculated from model output from the naive approach, and $U(\tilde N_{t1}^*), U(\tilde N_{t2}^*)$ denote Mann-Kendall statistics calculated from model output from the transfer learning approach for size class 1 and 2, respectively. Bold values indicate the best result between naive and transfer learning approaches for a given size class.}
    \centering
\begin{tabular}{lllcccc}
\toprule
Scenario & $\sigma^2_1$ & $\sigma^2_2$ & $U(\tilde N_{t1})$ & $U(\tilde N_{t2})$ & $U(\tilde N_{t1}^*)$ & $U(\tilde N_{t2}^*)$ \\
\midrule
I & 0.1 & 0.1 & 0.21 & 0.88 & \textbf{0.90} & \textbf{0.92}\\
II & 0.5 & 0.5 & 0.17 & 0.67 & \textbf{0.69} & \textbf{0.83}\\
III & 1.0 & 1.0 & 0.13 & 0.34 & \textbf{0.60} & \textbf{0.68}\\
IV & 0.2 & 0.8 & 0.18 & 0.38 & \textbf{0.78} & \textbf{0.68}\\
V & 0.8 & 0.2 & 0.20 & 0.79 & \textbf{0.78} & \textbf{0.94}\\
\bottomrule
\end{tabular}

    \label{tab:mk_cov}
\end{table}

\subsection{Potential for negative transfer under violated assumptions}

An important caveat of the transfer learning approach is that there is a (potentially) strong assumption that the source population (CR monitored population) and the target population (CPUE monitored population) are similar. Violations of this assumption can lead to negative transfer \citep{wang2019characterizing}, wherein inference on the target population is made worse by the transfer of information from a non-identical source population. In the previous simulation study, we assume that both the CR data and the CPUE data have the same detection function, i.e. the effect of covariates on detection is the same. Here we present two additional simulations where this assumption is violated to investigate the effects on inference. 

To test the efficacy of our method under slightly violated assumptions, we conduct two additional simulations under the setting with $\sigma^2_1=\sigma^2_2=0.1$ (scenario I). For both settings, the CR data are generated as above with $\boldsymbol\beta_1^{CR}=(-3.5,-2,0.5)'$. However, we generate CPUE data independently with (VI) $\boldsymbol\beta_1^{CPUE}=(-3.5,-1,0.5)'$ and (VII) $\boldsymbol\beta_1^{CPUE}=(-3.5,-3,0.5)'$ to analyze model performance when one of the covariate effects differs in magnitude, but is still significant and of the same sign.

\begin{table}[]
    \caption{Mean MAD of posterior mean from true population $\mathcal{N}_{tj}$ across simulations under each coefficient setting. $\tilde N_{t1}$ and $U(\tilde N_{t1})$ denotes abundance and Mann-Kendall statistic calculated from model output from the naive approach and $\tilde N_{t1}^*$ and $U(\tilde N_{t1}^*)$ denotes abundance and Mann-Kendall statistics calculated from model output from the transfer learning approach for size class 1. Bold values indicate the best result between naive and transfer learning approaches for a given size class.}
    \centering
    \begin{tabular}{llcccccc}
\toprule
&  & 
\multicolumn{2}{c}{MAD from $N_{tj}$} &
\multicolumn{2}{c}{Coverage of $N_{tj}$} &
\multicolumn{2}{c}{Coverage of $U(N_{tj})$} \\
\cmidrule(lr){3-4} \cmidrule(lr){5-6} \cmidrule(lr){7-8}
 Scenario & $\beta_1$ & $\tilde N_{t1}$ & $\tilde N_{t1}^*$ & $\tilde N_{t1}$ & $\tilde N_{t1}^*$ & $U(\tilde N_{t1})$ & $U(\tilde N_{t1}^*)$ \\
\midrule
VI   & -1 & 650.83 & \textbf{591.41} & 0.32 & \textbf{0.45} & \textbf{0.49} & 0.43 \\
VII  & -3 & 1596.59 & \textbf{634.33} & 0.11 & \textbf{0.45} & 0.07 & \textbf{0.35} \\
\bottomrule
\end{tabular}

    \label{tab:violated_sim}
\end{table}

The results of the simulations under scenarios (VI) and (VII) are reported in Table \ref{tab:violated_sim}. For these two simulations, we detect differences between the unadjusted and adjusted model estimates. Firstly, MAD improves under both scenarios using the adjustment, but significantly less so when $\beta_1=-1$. Similarly, the improvement in coverage for both abundance and trend detection is more remarkable for $\beta_1=-3$. This suggests that so long as the coefficient being transferred is of the same sign, then inference may be improved under transfer learning. Furthermore, transfer learning is more beneficial under the scenario where the effect is actually \textit{larger} in magnitude than we assumed. This follows from the idea that not accounting for a larger effect on detection probability (and therefore more variability in detection probability) would have more potential damage to ameliorate.

\section{Application to Recreationally Important Fisheries}\label{real_data_analysis}

\subsection{Data}\label{data}

\paragraph{Capture-recapture}  Smallmouth bass (\textit{Micropterus dolomieu}) were sampled following a capture-recapture procedure at a fixed-site on the Juniata River, PA from 1989-2001 and 2010-2013. Within each year, fish marking and recapture events were conducted over a 3-5 day period. Smallmouth bass were sampled using daytime boat electrofishing. Each fish, if not previously marked, was marked with a non-unique fin clip.

\paragraph{Catch per unit effort} Smallmouth bass CPUE data are more widely available throughout Pennsylvania. Across six large rivers, 19 river sections were sampled for CPUE. CPUE data for age-0 and older smallmouth bass were collected using using single-pass nighttime boat electrofishing at fixed sites and targeting similar seasonal conditions in which to survey, e.g. temperature, streamflow. The amount of time spent on a given sampling occasion ranged from less than one hour to six hours. Sampling was conducted by the Pennsylvania Fish and Boat Commission Fisheries Management Division from 1990 to 2022 and CPUE was recorded as the number of fish caught per hour of electrofishing. 

\paragraph{Covariates} We consider two sets of covariates in our model to capture the variation in abundance and detection. The abundance covariates, denoted $\textbf{z}_t$, is a vector consisting of an intercept term, year of sampling, and the average streamflow (cubic ft/s) in the 12 months preceding the sampling season (May-April) for each year $t$. Streamflow can influence smallmouth bass abundance both directly and indirectly by changing access to habitat and food resources, reducing reproductive success, and through direct mortality of eggs, fry, and juvenile fish \citep{lukas1995factors, freeman2022toward}. The detection covariates, $\textbf{x}_{tk}$, are a vector defined for each year $t$ and sampling day $k$. It consists of an intercept term, the hours of effort spent sampling, the daily streamflow, and a measure of temperature relative to the average temperature on that day of year across the whole time series, all of which have been shown to influence detection probability for freshwater fishes \citep{rosenberger2005validation, hangsleben2013evaluation, mcmanamay2014accounting, gwinn2016imperfect}. 

\subsection{Analysis}

We first analyzed the capture-recapture dataset. Then, the posterior draws of the detection function parameters were transferred to the CPUE model for subsequent analysis. For our study, we consider two size classes of smallmouth bass that are ecologically meaningful --  those greater than and less than 200 mm in length. In our study systems, fish less than 200 mm in length are generally immature and fish greater than 200 mm in length are mature (adult) fish. 

\subsubsection{CR analysis}

The analysis of the smallmouth bass CR data provides not only the necessary pieces to augment our CPUE analysis, but valuable insights in its own right. Table \ref{tab:real_betas} shows the mean and credible intervals for the coefficients of the detection covariates for the two size classes. Of particular interest are estimates of $\beta_2$ and $\beta_3$, which capture the effect of streamflow and relative temperature on detection, respectively. We note that streamflow has a larger negative and significant effect on detection for smaller fish than larger fish, based on 95\% credible intervals not containing 0, suggesting that higher daily streamflow reduce the detection probability of smaller fish. Similarly, relative temperature has a positive effect on the detection of smaller fish and no significant effect on larger fish, which indicates that warmer days correspond to higher detection probabilities for smaller individuals.

Overall, detection probabilities vary across size classes not only in covariate effects, but also in absolute probabilities. The intercept term, $\beta_0$, is greater for the larger ($>200$ mm) fish, meaning an overall higher average detection rate, which can be visualized in Figure S1 found in the Appendix S1. The significant difference in detection functions and probabilities between the two size classes indicates that a successful transfer to a CPUE model requires this delineation in size classes, and is a good example of how a poor assumption (i.e., in this case that detection for both fish size classes respond similarly to the environment) has the potential for misleading inference.

\begin{table}[ht]
\caption{Posterior means and 95\% credible intervals for fixed effects on detection probability for two size classes of smallmouth bass (\textit{Micropterus dolomieu}) sampled from the Juniata River, PA. $\beta_0$ = intercept; $\beta_1$ = effort (hrs); $\beta_2$ = streamflow (cfs), and $\beta_3$ = relative temperature ($^\circ$C).}
\centering
\begin{tabular}{lcc lcc}
\toprule
& \multicolumn{2}{c}{$<200$ mm} & & \multicolumn{2}{c}{$>200$ mm} \\
\cmidrule{2-3} \cmidrule{5-6}
 & Mean & 95\% Credible Interval & & Mean & 95\% Credible Interval \\
\midrule
$\beta_0$ & -4.30 & (-4.85, -3.78) & & -2.56 & (-3.39, -1.74) \\
$\beta_1$ &  0.13 & (0.00, 0.26)      & &  0.26 & (0.04, 0.47) \\
$\beta_2$ & -1.92 & (-2.71, -1.29)  & & -0.87 & (-1.77, 0.01) \\
$\beta_3$ &  0.74 & (0.26, 1.22)   & &  0.01 & (-0.78, 0.80) \\
\bottomrule
\end{tabular}

\label{tab:real_betas}
\end{table}

In addition to learning about detection probabilities, a more common inferential goal in CR studies is to learn true abundance. Figure S2 shows abundance estimates across the time period of study, highlighting expected significant differences between and across size classes and years. The smaller ($<$200 mm) fish appear to have declined in abundance over the study duration, while the larger ($>$200 mm) fish have remained somewhat constant with a possible increase in the later years.

\subsubsection{Transfer to CPUE Data}

CPUE data were collected from 19 different stream locations across six Pennsylvania rivers. For each location, we used the naive approach (\ref{cpue_model}) and the transfer learning approach (\ref{adjusted_model}), which compensates for variable detection probabilities via transfer learning. Here, the adjustment is obtained by inserting posterior samples of $\pi(\beta_j|(\textbf{n}-\textbf{m},\textbf{m}))$ for $j=0,2,3$ learned from the Juniata CR data analysis above. In this subsequent model, we do not include $\beta_1$, which corresponds to the effect of effort in the CR data because CPUE already takes effort into account in its formulation. Here we discuss the results for four different stream segments (only showing two graphically to conserve space), but the results of all populations are available in Appendix S1. First, we consider two segments of the Juniata River (referred to as Juniata River segments 3 and 4), which are distinct from the one where the CR data were collected. We also consider segments of the Susquehanna River, PA (referred to as Susquehanna River segments 5 and 6).

\begin{figure}
    \centering
    \includegraphics[width=0.8\linewidth]{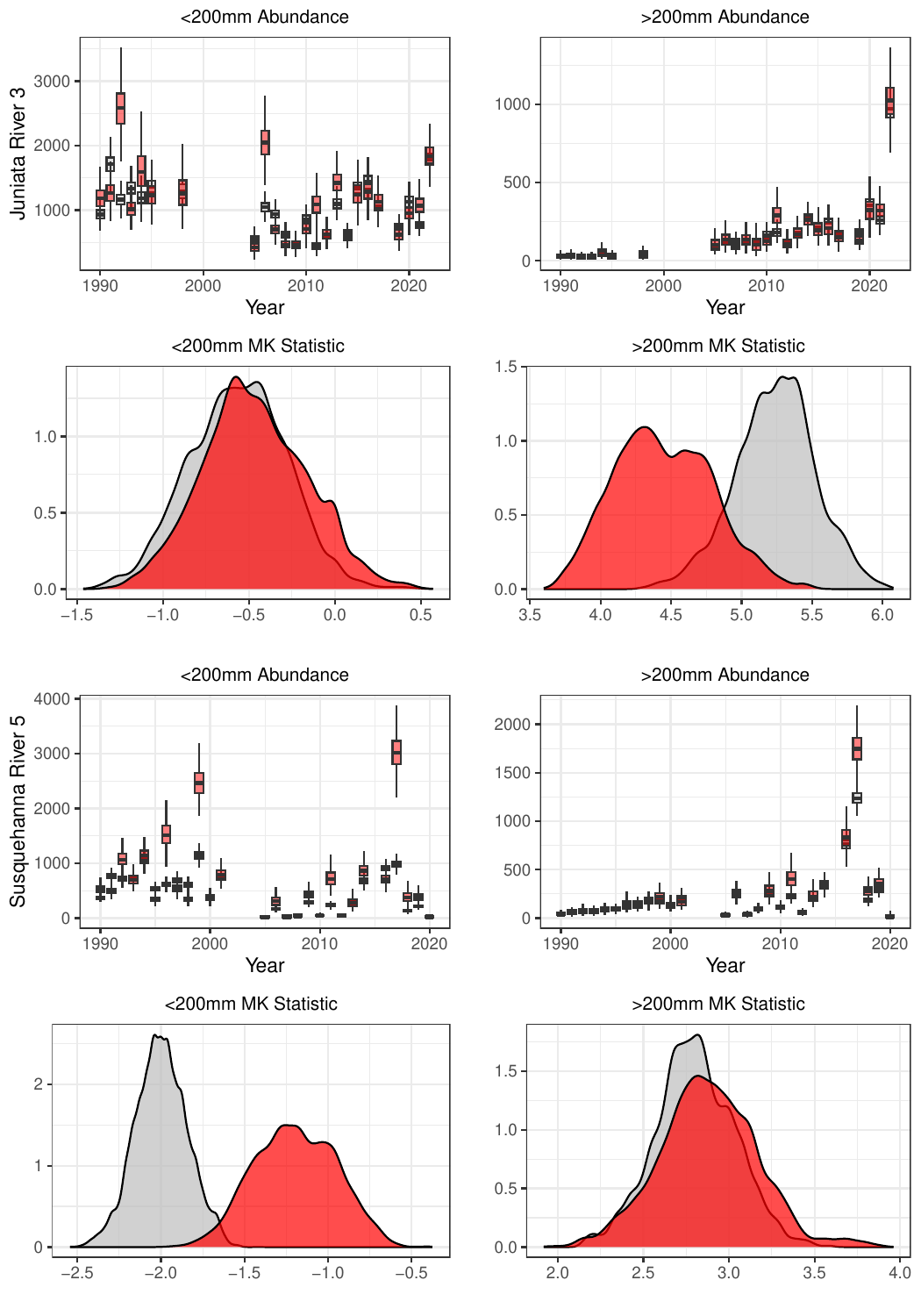}
    \caption{Transfer learning model output for Juniata River segment 3 and Susquehanna River segment 5 CPUE data for smallmouth bass (\textit{Micropterus dolomieu}) sampled in the Juniata and Susquehanna Rivers, PA. The boxplots in rows one and three are posterior distributions for abundance from the naive approach ($\tilde N_{tj}$; grey) and transfer learning approach ($\tilde N_{tj}^*$; red). Rows two and four are posterior distributions for the naive approach ($U(\tilde N_{tj})$; grey) and transfer learning approach ($U(\tilde N_{tj}^*)$; red).}
    \label{fig:cpue_results}
\end{figure}

Posterior distributions of population estimates and trend statistics for Juniata River segment 3 are shown in Figure \ref{fig:cpue_results} (top four panels). The largest differences between the two approaches are detected for the $<200$ mm size class, where several years have significantly different abundance estimates. For the larger size class, all of the yearly population estimates are generally similar. With respect to trend detection, the Mann-Kendall statistic of $\tilde N_{t1}^*$ is similar in expectation to $\tilde N_{t1}$, but has a higher variance resulting in a more significant overlap of 0, i.e., less evidence of a trend. For the larger size class, $U(\tilde N_{t2}^*)$ is again more diffuse than $U(\tilde N_{t2})$, but the mean shifts dramatically towards 0, indicating a less positive trend than found in the naive approach.

Similar results are seen for Juniata River segment 4 (Figure S4). In general, $\tilde N_{t1}^*$ shows greater deviation from $\tilde N_{t1}$ than $\tilde N_{t2}^*$ does from $\tilde N_{t2}$. The statistic $U(\tilde N_{tj}^*)$ is once again more diffuse than its unadjusted counterpart for both size classes. Interestingly, $U(\tilde N_{t1}^*)$ has a larger expectation than $U(\tilde N_{t1})$, but this is counteracted by a credible interval that includes 0. The differences between $U(\tilde N_{t2}^*)$ and $U(\tilde N_{t2})$ mirror those in segment 3.

For Susquehanna River segment 5, there are many significant differences in population estimates for the smaller fish between the naive and transfer learning approaches (Figure \ref{fig:cpue_results}, bottom four panels). Additionally, for the larger fish, there are several instances where $\tilde N_{t2}$ is significantly larger than $\tilde N_{t2}$. We see a similar effect as in the Juniata analyses where the Mann-Kendall posteriors are more diffuse under the transfer learning approach. Notably, $U(\tilde N_{t1}^*)$ has a less negative expectation than $U(\tilde N_{t1})$. This is particularly notable because smallmouth bass experienced declines in overall population in the mid 2000's due to a disease outbreak \citep{schall2020fishery}. While the transfer learning approach still indicates a decline in population, it is less severe than the one estimated using the naive approach. Susquehanna River segment 6 has similar discrepancies throughout the time period of analysis between $\tilde N_{tj}$ and $\tilde N_{tj}^*$; however, this is not reflected in a significantly shifted mean for $U(\tilde N_{tj}^*)$ (Figure S19). We once again see a more diffuse posterior indicating that the naive approach is overly confident in its trend assessment.

\section{Discussion}

We have developed a methodological framework for transferring information on detection functions learned via capture-recapture studies into analyses of catch per unit effort data to improve inference on both absolute abundance and temporal trends. It was demonstrated empirically that our Bayesian transfer learning approach improved inference with respect to both goals, and injected additional uncertainty that improved coverage of the resulting posterior credible intervals. Of particular interest in the case where CR and CPUE data may be collected from distinct populations, we showed that our method is robust to minor violations of the transferability assumption. That is, even if the detection functions are not identical across populations, so long as they are directionally similar, inference can still benefit from transfer learning.

We applied our method to data collected of smallmouth bass from river systems in Pennsylvania, USA. First, results from a hierarchical Bayesian analysis of capture-recapture data in the Juniata River reveal significant effects of both daily streamflow and daily relative temperature on detection probability. Furthermore, we find that these effects and the resulting detection probabilities differ between size classes (juvenile vs. adult fish) of smallmouth bass. With these learned detection functions, we transfer information into our CPUE models of smallmouth bass populations in distinct river segments in Pennsylvania. One notable result came from the Susquehanna River where the smallmouth bass population has been documented to experience a recent disease outbreak that negatively impacted population. We found that under our adjustment, the population decrease was still present, but perhaps less severe than initially estimated based on pure count data.

Although we have demonstrated that transfer learning in the context of CR/CPUE can be helpful, the assumption of transferability is strong and may be a limitation in some inferential contexts. We illustrate how minor violations of this assumption can still lead to improved inference, but further research is needed to understand the limits of these acceptable violations. Furthermore, an important body of work in transfer learning \citep{wang2019characterizing} focuses on limiting negative transfer, which is challenging in this case, as the count data offers no information as to appropriate detection functions. However, this problem could be mitigated with domain expertise that provides a richer understanding of how detection functions vary across factors such as species, habitat conditions, and capture techniques. For example, in the context of fisheries, it is well established that streamflow tends to have a negative effect on detection probability \citep{white2020predicting}, making the assumption of transferability with only mild heterogeneity reasonable in this setting. This is also true for many terrestrial systems, where it can be assumed that the direction of the effect (positive or negative) of a catchability/detection covariate does not change, although the magnitude of the effect does \citep[e.g., the negative effect of wind and the positive effect of observer experience on detection probability during bird point counts; ][]{rigby2019factors, bergen2023effects}.

\section*{Acknowledgments}
This research was supported by the U.S. Geological Survey Northeast and Southeast Climate Adaptation Science Center Grant No. G22AC00597-01. We would like to thank Pennsylvania Fish \& Boat Commission (PFBC) biologists and staff for collection of field data during smallmouth bass population surveys funded by the Federal Aid in Sportfish Restoration (F-57-R). We thank Bob Lorantas and Geoff Smith for providing information about PFBC sampling methodology. Any use of trade, firm, or product names is for descriptive purposes only and does not imply endorsement by the U.S. Government. The code for model fitting is available at https://doi.org/10.5066/P1XJ5UDV. The data underlying this article were provided by the Pennsylvania Fish \& Boat Commission by permission. Data can be obtained by request to G.D. Smith geofsmith@pa.gov.


%
\bibliography{refs.bib}

@article{rigby2019factors,
  title={Factors affecting detection probability, effective area surveyed, and species misidentification in grassland bird point counts},
  author={Rigby, Elizabeth A and Johnson, Douglas H},
  journal={The Condor},
  volume={121},
  number={3},
  pages={duz030},
  year={2019},
  publisher={Oxford University Press US}
}

@article{bergen2023effects,
  title={Effects of observer skill and survey method on forest bird abundance data: Recommendations for citizen science conservation monitoring in the Caribbean},
  author={Bergen, Nicholas and De Ruyck, Christopher C and Koper, Nicola and others},
  journal={Journal of Caribbean Ornithology},
  volume={36},
  pages={45--61},
  year={2023}
}

@article{white2020predicting,
  title={Predicting fish species richness and habitat relationships using Bayesian hierarchical multispecies occupancy models},
  author={White, Shannon and Faulk, Evan and Tzilkowski, Caleb and Weber, Andrew and Marshall, Matthew and Wagner, Tyler},
  journal={Canadian Journal of Fisheries and Aquatic Sciences},
  volume={77},
  number={3},
  pages={602--610},
  year={2020},
  publisher={NRC Research Press}
}

@article{hangsleben2013evaluation,
  title={Evaluation of electrofishing catch per unit effort for indexing fish abundance in Florida lakes},
  author={Hangsleben, Matt A and Allen, Micheal S and Gwinn, Daniel C},
  journal={Transactions of the American Fisheries Society},
  volume={142},
  number={1},
  pages={247--256},
  year={2013},
  publisher={Oxford University Press Oxford, UK}
}

@article{mcmanamay2014accounting,
  title={Accounting for variation in species detection in fish community monitoring},
  author={McManamay, RA and Orth, DJ and Jager, HI},
  journal={Fisheries Management and Ecology},
  volume={21},
  number={2},
  pages={96--112},
  year={2014},
  publisher={Wiley Online Library}
}

@article{rosenberger2005validation,
  title={Validation of abundance estimates from mark--recapture and removal techniques for rainbow trout captured by electrofishing in small streams},
  author={Rosenberger, Amanda E and Dunham, Jason B},
  journal={North American Journal of Fisheries Management},
  volume={25},
  number={4},
  pages={1395--1410},
  year={2005},
  publisher={Oxford University Press Oxford, UK}
}

@article{gwinn2016imperfect,
  title={Imperfect detection and the determination of environmental flows for fish: challenges, implications and solutions},
  author={Gwinn, Daniel C and Beesley, Leah S and Close, Paul and Gawne, Ben and Davies, Peter M},
  journal={Freshwater Biology},
  volume={61},
  number={1},
  pages={172--180},
  year={2016},
  publisher={Wiley Online Library}
}

@article{freeman2022toward,
  title={Toward improved understanding of streamflow effects on freshwater fishes},
  author={Freeman, Mary C and Bestgen, Kevin R and Carlisle, Daren and Frimpong, Emmanuel A and Franssen, Nathan R and Gido, Keith B and Irwin, Elise and Kanno, Yoichiro and Luce, Charles and Kyle McKay, S and others},
  journal={Fisheries},
  volume={47},
  number={7},
  pages={290--298},
  year={2022},
  publisher={Oxford University Press Oxford, UK}
}

@article{lukas1995factors,
  title={Factors affecting nesting success of smallmouth bass in a regulated Virginia stream},
  author={Lukas, Joseph A and Orth, Donald J},
  journal={Transactions of the American Fisheries Society},
  volume={124},
  number={5},
  pages={726--735},
  year={1995},
  publisher={Taylor \& Francis}
}

@article{ye2009reliable,
  title={How reliable are the abundance indices derived from commercial catch--effort standardization?},
  author={Ye, Yimin and Dennis, Darren},
  journal={Canadian Journal of Fisheries and Aquatic Sciences},
  volume={66},
  number={7},
  pages={1169--1178},
  year={2009}
}

@article{ducharme2018indices,
  title={Indices of abundance in the Gulf of Mexico reef fish complex: A comparative approach using spatial data from vessel monitoring systems},
  author={Ducharme-Barth, Nicholas D and Shertzer, Kyle W and Ahrens, Robert NM},
  journal={Fisheries Research},
  volume={198},
  pages={1--13},
  year={2018},
  publisher={Elsevier}
}

@article{hooten2007hierarchical,
  title={Hierarchical spatiotemporal matrix models for characterizing invasions},
  author={Hooten, Mevin B and Wikle, Christopher K and Dorazio, Robert M and Royle, J Andrew},
  journal={Biometrics},
  volume={63},
  number={2},
  pages={558--567},
  year={2007},
  publisher={Oxford University Press}
}

@article{murray2020trends,
  title={Trends in inland commercial fisheries in the United States},
  author={Murray, Devin N and Bunnell, David B and Rogers, Mark W and Lynch, Abigail J and Douglas Beard Jr, T and Funge-Smith, Simon},
  journal={Fisheries},
  volume={45},
  number={11},
  pages={585--596},
  year={2020},
  publisher={Oxford University Press Oxford, UK}
}

@article{funge2019fresh,
  title={A fresh look at inland fisheries and their role in food security and livelihoods},
  author={Funge-Smith, Simon and Bennett, Abigail},
  journal={Fish and Fisheries},
  volume={20},
  number={6},
  pages={1176--1195},
  year={2019},
  publisher={Wiley Online Library}
}

@article{seber1965note,
  title={A note on the multiple-recapture census},
  author={Seber, George AF},
  journal={Biometrika},
  volume={52},
  number={1/2},
  pages={249--259},
  year={1965},
  publisher={JSTOR}
}

@article{jolly1982mark,
  title={Mark-recapture models with parameters constant in time},
  author={Jolly, GM},
  journal={Biometrics},
  pages={301--321},
  year={1982},
  publisher={JSTOR}
}

@article{staton2022accounting,
  title={Accounting for uncertainty when estimating drivers of imperfect detection: An integrated approach illustrated with snorkel surveys for riverine fishes},
  author={Staton, Benjamin A and Justice, Casey and White, Seth and Sedell, Edwin R and Burns, Lauren A and Kaylor, Matthew J},
  journal={Fisheries Research},
  volume={249},
  pages={106209},
  year={2022},
  publisher={Elsevier}
}

@article{kellner2014accounting,
  title={Accounting for imperfect detection in ecology: a quantitative review},
  author={Kellner, Kenneth F and Swihart, Robert K},
  journal={PloS one},
  volume={9},
  number={10},
  pages={e111436},
  year={2014},
  publisher={Public Library of Science San Francisco, USA}
}

@article{allen2015evaluating,
  title={Evaluating and validating abundance monitoring methods in the absence of populations of known size: review and application to a passive tracking index},
  author={Allen, Lee R and Engeman, Richard M},
  journal={Environmental Science and Pollution Research},
  volume={22},
  number={4},
  pages={2907--2915},
  year={2015},
  publisher={Springer}
}

@article{pollock2002large,
  title={Large scale wildlife monitoring studies: statistical methods for design and analysis},
  author={Pollock, Kenneth H and Nichols, James D and Simons, Theodore R and Farnsworth, George L and Bailey, Larissa L and Sauer, John R},
  journal={Environmetrics: The official journal of the International Environmetrics Society},
  volume={13},
  number={2},
  pages={105--119},
  year={2002},
  publisher={Wiley Online Library}
}

@article{schall2020fishery,
  title={A fishery after the decline: the Susquehanna River Smallmouth Bass story},
  author={Schall, Megan K and Smith, Geoffrey D and Blazer, Vicki S and Walsh, Heather L and Li, Yan and Wagner, Tyler},
  journal={Fisheries},
  volume={45},
  number={11},
  pages={576--584},
  year={2020},
  publisher={Oxford University Press Oxford, UK}
}

@article{hubert2007relative,
  title={Relative abundance and catch per unit effort},
  author={Hubert, Wayne A and Fabrizio, Mary C},
  journal={Analysis and interpretation of freshwater fisheries data. American Fisheries Society, Bethesda, Maryland},
  pages={279--325},
  year={2007}
}

@article{castledine1981bayesian,
  title={A Bayesian analysis of multiple-recapture sampling for a closed population},
  author={Castledine, BJ},
  journal={Biometrika},
  volume={68},
  number={1},
  pages={197--210},
  year={1981},
  publisher={Oxford University Press}
}

@article{ver2025integrated,
  title={An integrated data model to estimate abundance from counts with temporal dependence and imperfect detection},
  author={Ver Hoef, Jay M and McClintock, Brett T and Boveng, Peter L and London, Josh M and Jansen, John K},
  journal={Ecology},
  volume={106},
  number={5},
  pages={e70073},
  year={2025},
  publisher={Wiley Online Library}
}

@article{north2024accounting,
  title={Accounting for spatiotemporal sampling variation in joint species distribution models},
  author={North, Joshua S and Schliep, Erin M and Hansen, Gretchen JA and Kundel, Holly and Custer, Christopher A and McLaughlin, Paul and Wagner, Tyler},
  journal={Journal of Applied Ecology},
  volume={61},
  number={1},
  pages={186--201},
  year={2024},
  publisher={Wiley Online Library}
}

@article{weiss2016survey,
  title={A survey of transfer learning},
  author={Weiss, Karl and Khoshgoftaar, Taghi M and Wang, DingDing},
  journal={Journal of Big data},
  volume={3},
  number={1},
  pages={9},
  year={2016},
  publisher={Springer}
}

@article{suder2023bayesian,
  title={Bayesian transfer learning},
  author={Suder, Piotr M and Xu, Jason and Dunson, David B},
  journal={arXiv preprint arXiv:2312.13484},
  year={2023}
}

@article{fieberg2012estimating,
  title={Estimating population abundance using sightability models: R SightabilityModel package},
  author={Fieberg, John R},
  journal={Journal of Statistical Software},
  volume={51},
  pages={1--20},
  year={2012}
}

@inproceedings{wang2019characterizing,
  title={Characterizing and avoiding negative transfer},
  author={Wang, Zirui and Dai, Zihang and P{\'o}czos, Barnab{\'a}s and Carbonell, Jaime},
  booktitle={Proceedings of the IEEE/CVF conference on computer vision and pattern recognition},
  pages={11293--11302},
  year={2019}
}

@book{ovaskainen2020joint,
  title={Joint species distribution modelling: With applications in R},
  author={Ovaskainen, Otso and Abrego, Nerea},
  year={2020},
  publisher={Cambridge University Press}
}

@article{kendall1948rank,
  title={Rank correlation methods.},
  author={Kendall, Maurice George},
  year={1948},
  publisher={Griffin}
}



\label{lastpage}

\newpage
\section{Appendix S1}

\begin{figure}[H]
    \centering
    \includegraphics[width=0.8\linewidth]{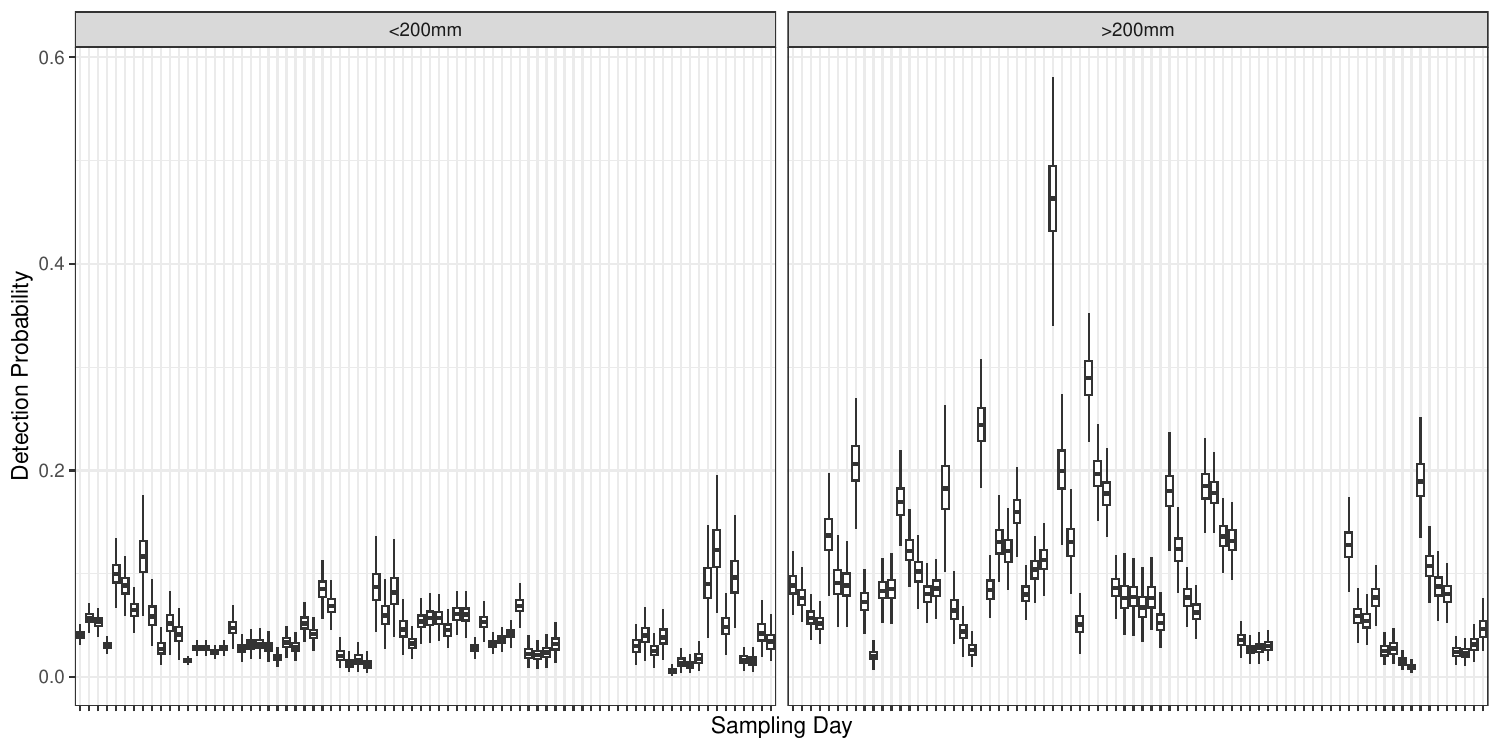}
    \caption{Posterior distributions of daily detection probabilities ($p_{tjk}$) for two size classes of riverine smallmouth bass (\textit{Micropterus dolomieu}) sampled in the Juniata River, PA.}
    \label{fig:CR_p_box}
\end{figure}

\begin{figure}[H]
    \centering
    \includegraphics[width=0.8\linewidth]{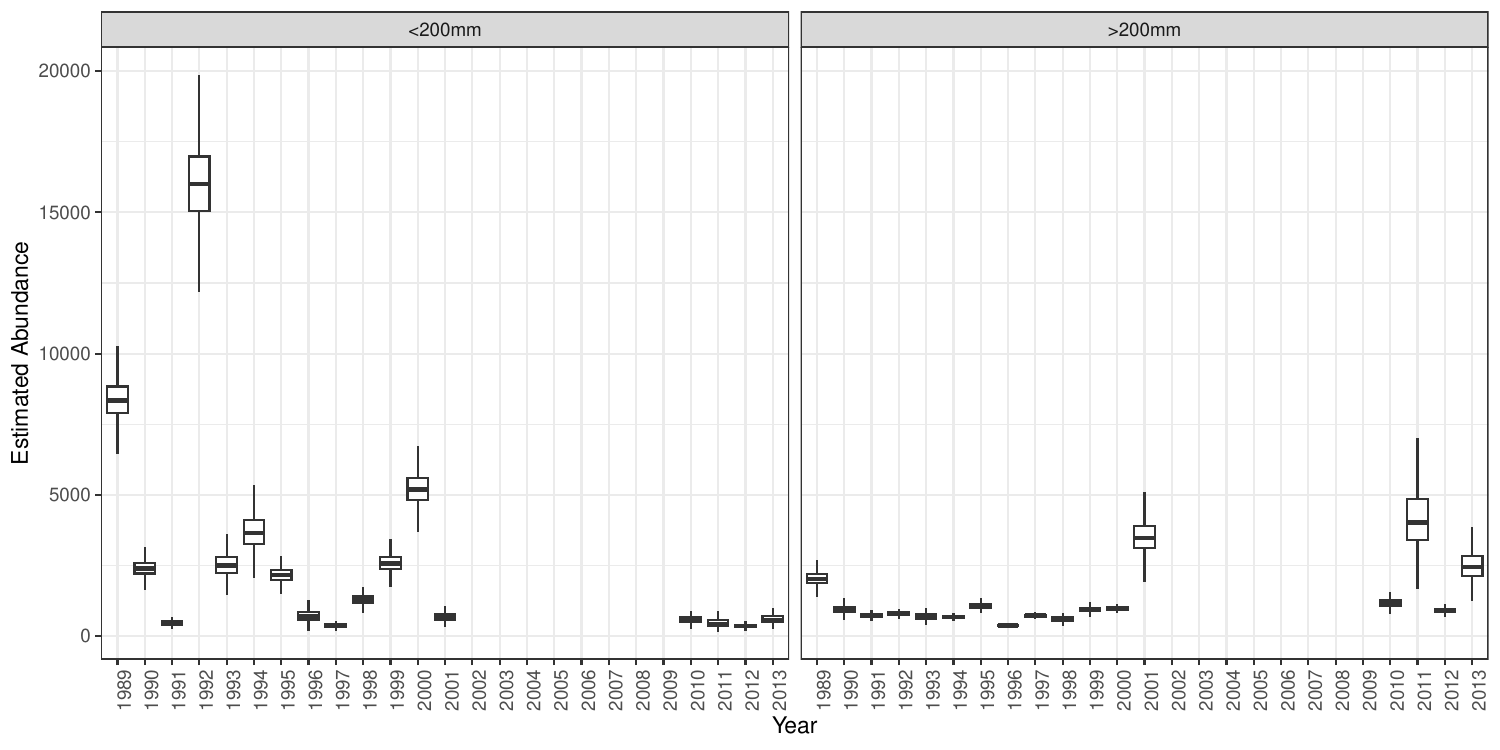}
    \caption{Posterior distributions of annual abundance ($N_{tj}$) for two size classes of riverine smallmouth bass (\textit{Micropterus dolomieu}) sampled in the Juniata River, PA.}
    \label{fig:CR_N_box}
\end{figure}

\begin{figure}
    \centering
    \includegraphics[width=0.8\linewidth]{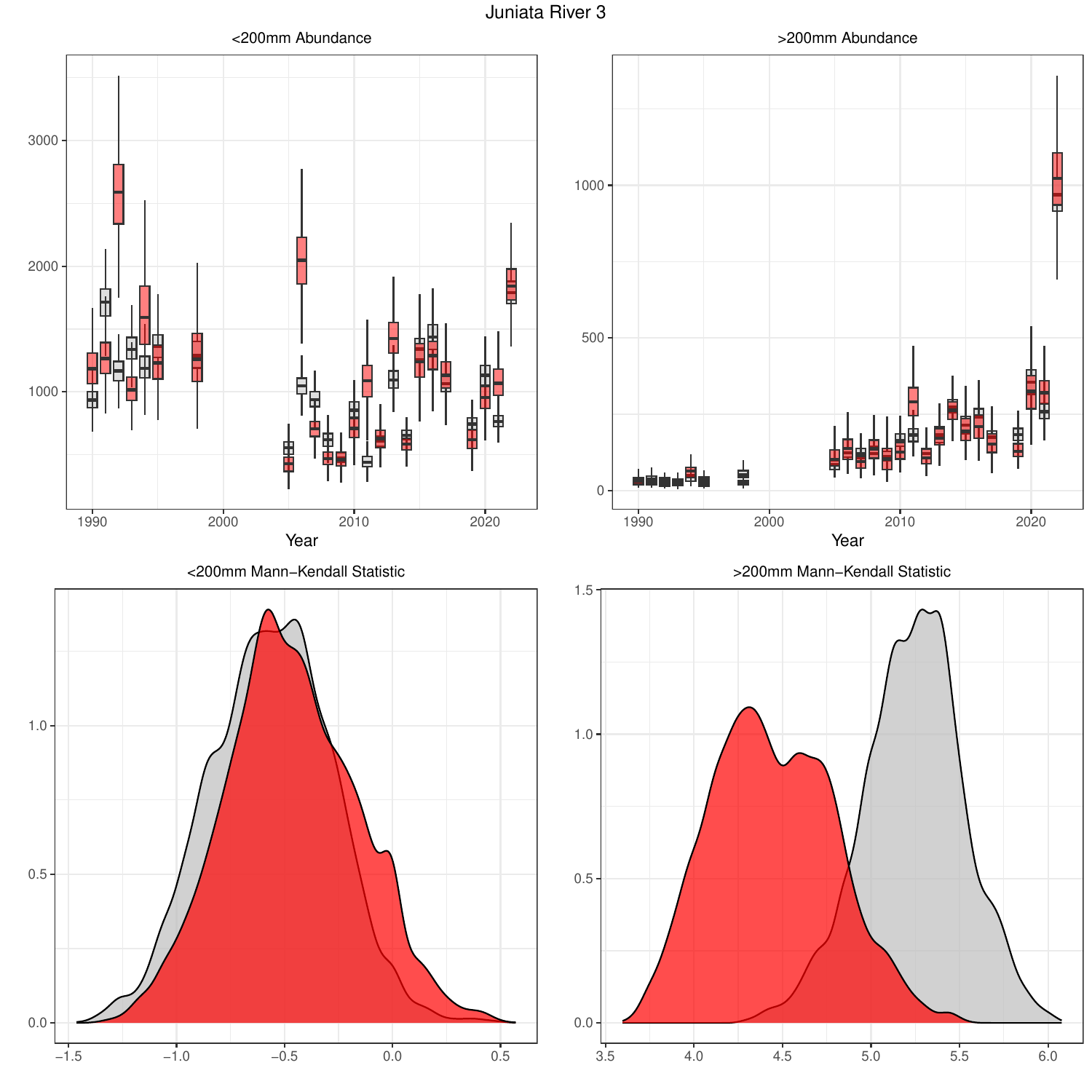}
    \caption{Model output for Juniata River Segment 3 CPUE data. The top two panels are posterior distributions for $\tilde N_{tj}$ (grey) and $\tilde N_{tj}^*$ (red). The bottom two panels are posterior distributions for $U(\tilde N_{tj})$ (grey) and $U(\tilde N_{tj}^*)$ (red).}
    \label{fig:jun3}
\end{figure}

\begin{figure}
    \centering
    \includegraphics[width=0.8\linewidth]{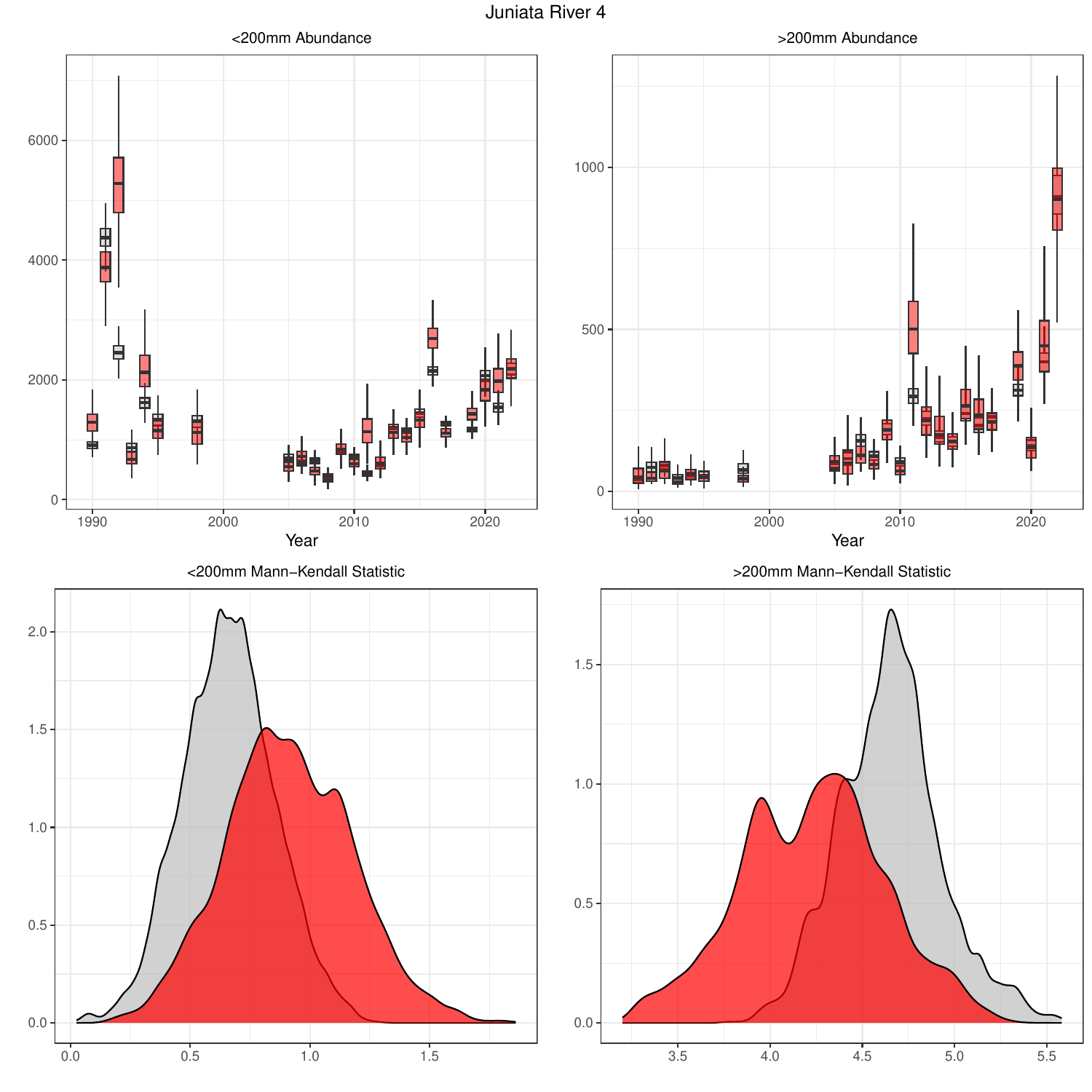}
    \caption{Model output for Juniata River Segment 4 CPUE data. The top two panels are posterior distributions for $\tilde N_{tj}$ (grey) and $\tilde N_{tj}^*$ (red). The bottom two panels are posterior distributions for $U(\tilde N_{tj})$ (grey) and $U(\tilde N_{tj}^*)$ (red).}
    \label{fig:jun4}
\end{figure}

\begin{figure}
    \centering
    \includegraphics[width=0.8\linewidth]{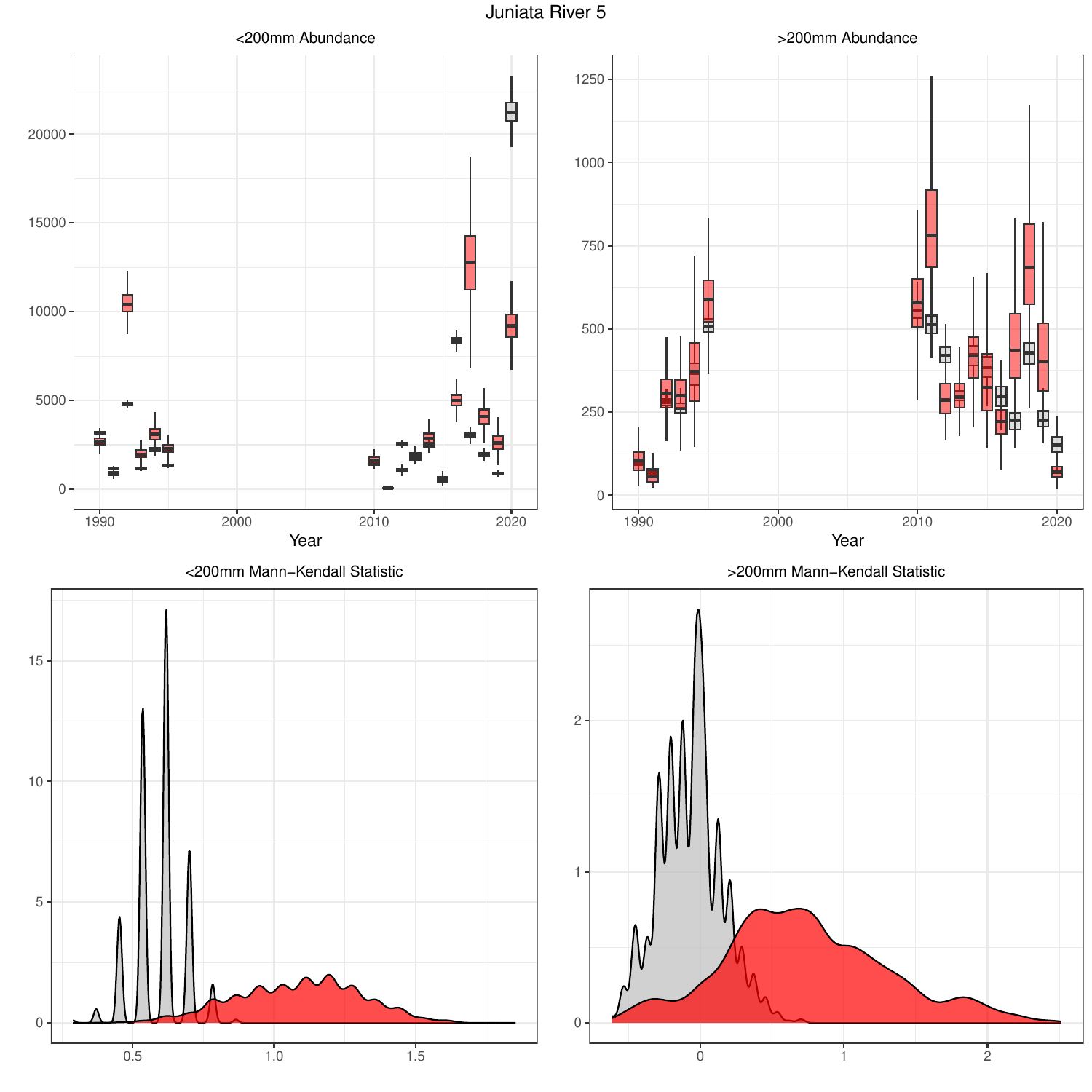}
    \caption{Model output for Juniata River Segment 5 CPUE data. The top two panels are posterior distributions for $\tilde N_{tj}$ (grey) and $\tilde N_{tj}^*$ (red). The bottom two panels are posterior distributions for $U(\tilde N_{tj})$ (grey) and $U(\tilde N_{tj}^*)$ (red).}
    \label{fig:jun5}
\end{figure}

\begin{figure}
    \centering
    \includegraphics[width=0.8\linewidth]{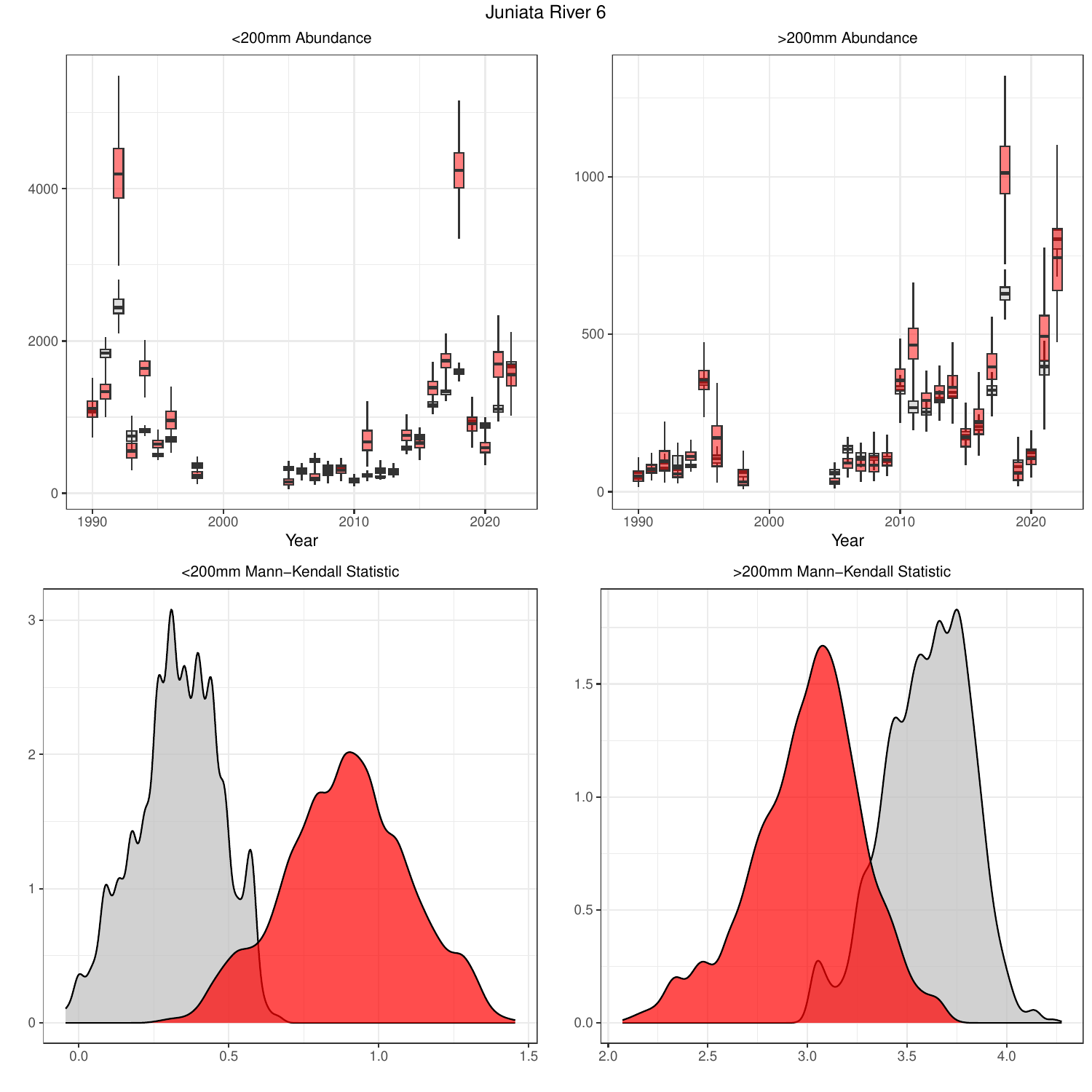}
    \caption{Model output for Juniata River Segment 6 CPUE data. The top two panels are posterior distributions for $\tilde N_{tj}$ (grey) and $\tilde N_{tj}^*$ (red). The bottom two panels are posterior distributions for $U(\tilde N_{tj})$ (grey) and $U(\tilde N_{tj}^*)$ (red).}
    \label{fig:jun6}
\end{figure}

\begin{figure}
    \centering
    \includegraphics[width=0.8\linewidth]{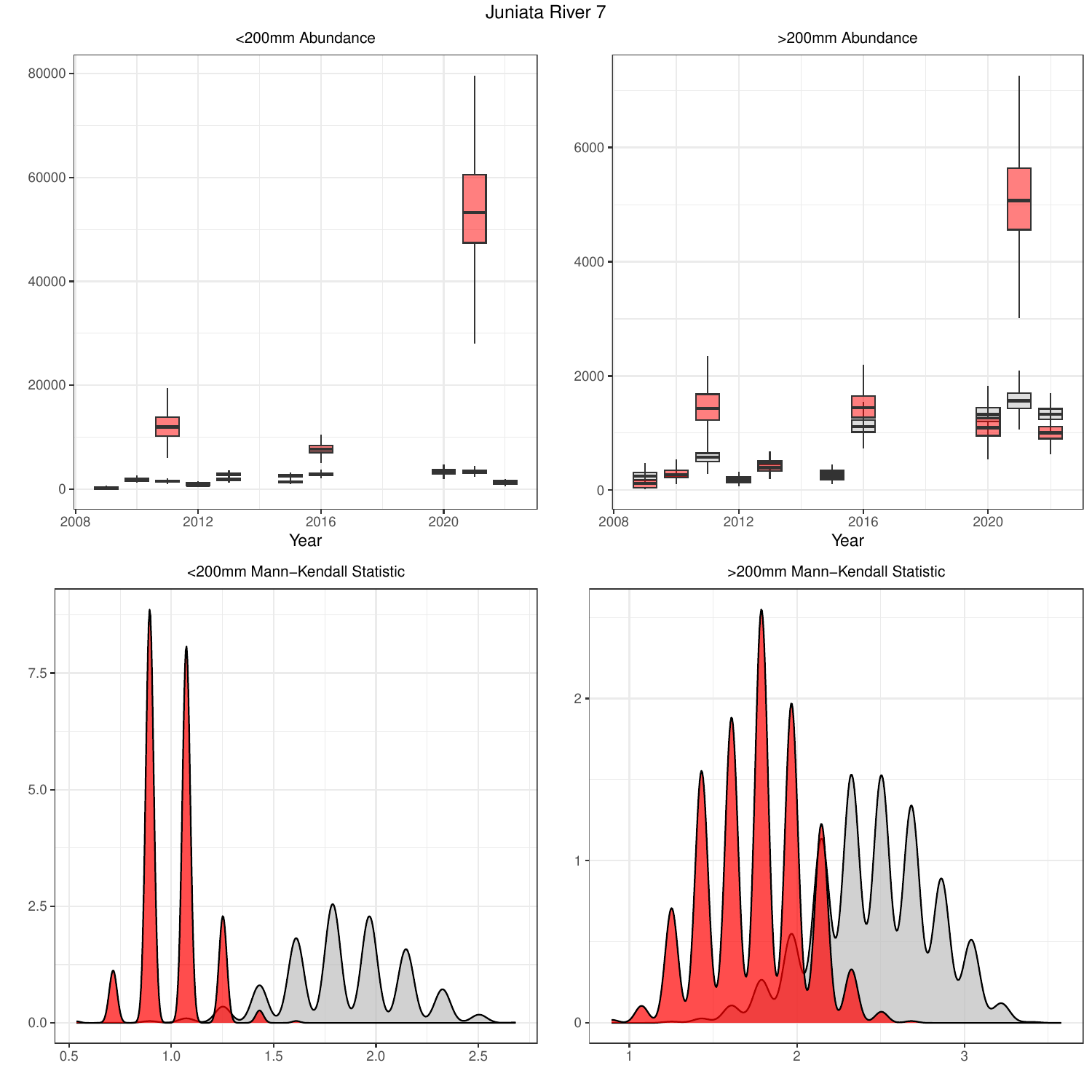}
    \caption{Model output for Juniata River Segment 7 CPUE data. The top two panels are posterior distributions for $\tilde N_{tj}$ (grey) and $\tilde N_{tj}^*$ (red). The bottom two panels are posterior distributions for $U(\tilde N_{tj})$ (grey) and $U(\tilde N_{tj}^*)$ (red).}
    \label{fig:jun7}
\end{figure}

\begin{figure}
    \centering
    \includegraphics[width=0.8\linewidth]{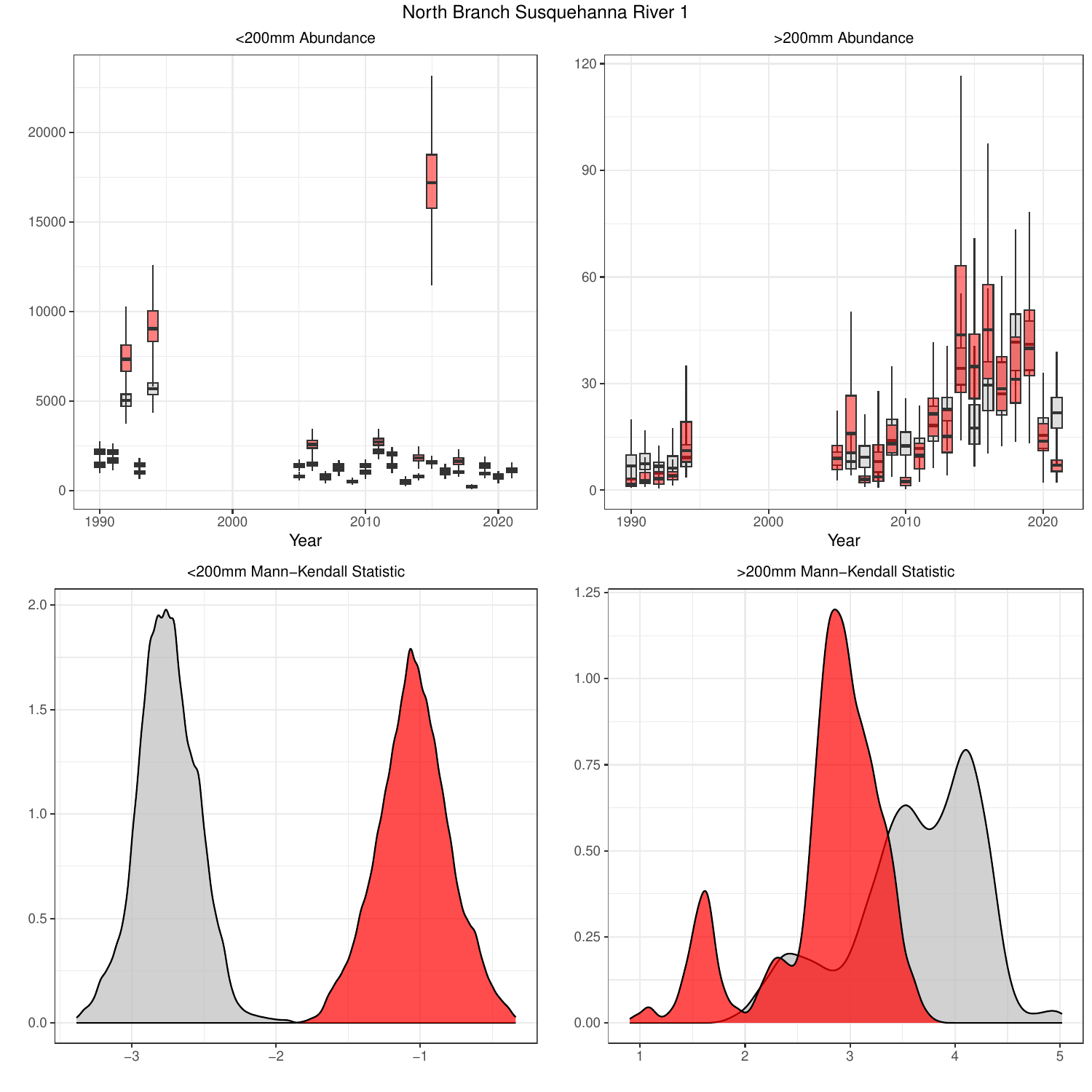}
    \caption{Model output for North Branch Susquehanna River Segment 1 CPUE data. The top two panels are posterior distributions for $\tilde N_{tj}$ (grey) and $\tilde N_{tj}^*$ (red). The bottom two panels are posterior distributions for $U(\tilde N_{tj})$ (grey) and $U(\tilde N_{tj}^*)$ (red).}
    \label{fig:nbsusq1}
\end{figure}

\begin{figure}
    \centering
    \includegraphics[width=0.8\linewidth]{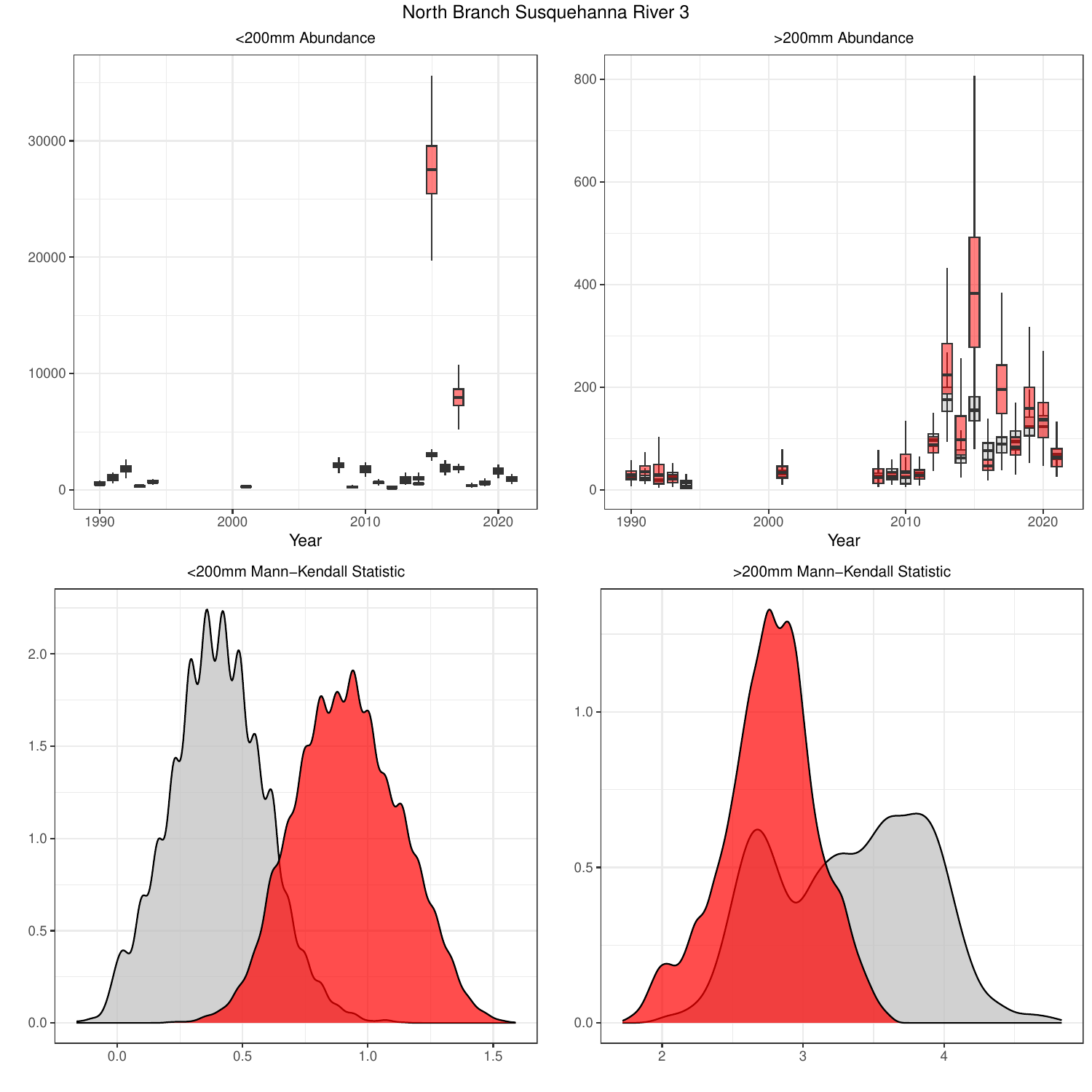}
    \caption{Model output for North Branch Susquehanna River Segment 3 CPUE data. The top two panels are posterior distributions for $\tilde N_{tj}$ (grey) and $\tilde N_{tj}^*$ (red). The bottom two panels are posterior distributions for $U(\tilde N_{tj})$ (grey) and $U(\tilde N_{tj}^*)$ (red).}
    \label{fig:nbsusq3}
\end{figure}

\begin{figure}
    \centering
    \includegraphics[width=0.8\linewidth]{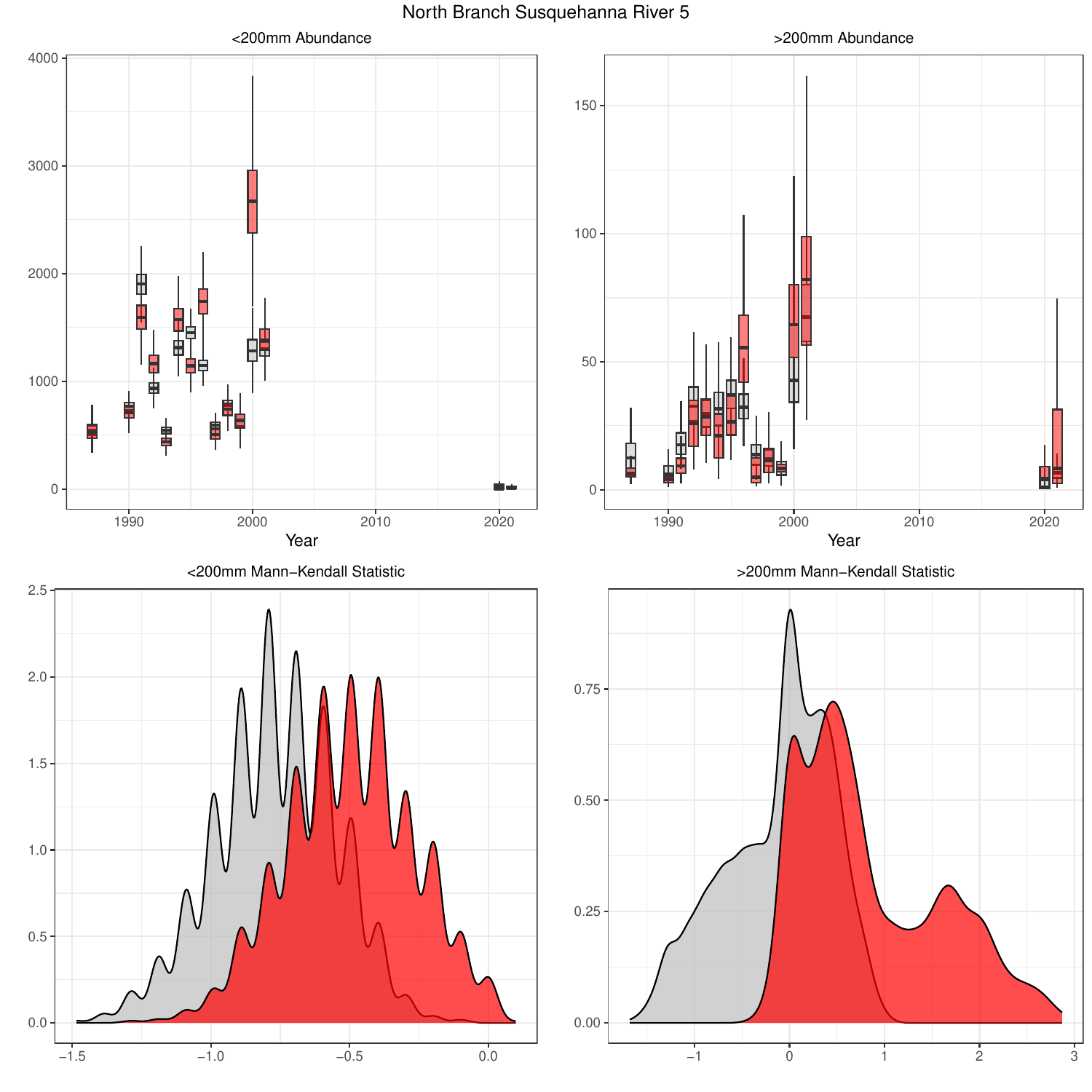}
    \caption{Model output for North Branch Susquehanna River Segment 5 CPUE data. The top two panels are posterior distributions for $\tilde N_{tj}$ (grey) and $\tilde N_{tj}^*$ (red). The bottom two panels are posterior distributions for $U(\tilde N_{tj})$ (grey) and $U(\tilde N_{tj}^*)$ (red).}
    \label{fig:nbsusq5}
\end{figure}

\begin{figure}
    \centering
    \includegraphics[width=0.8\linewidth]{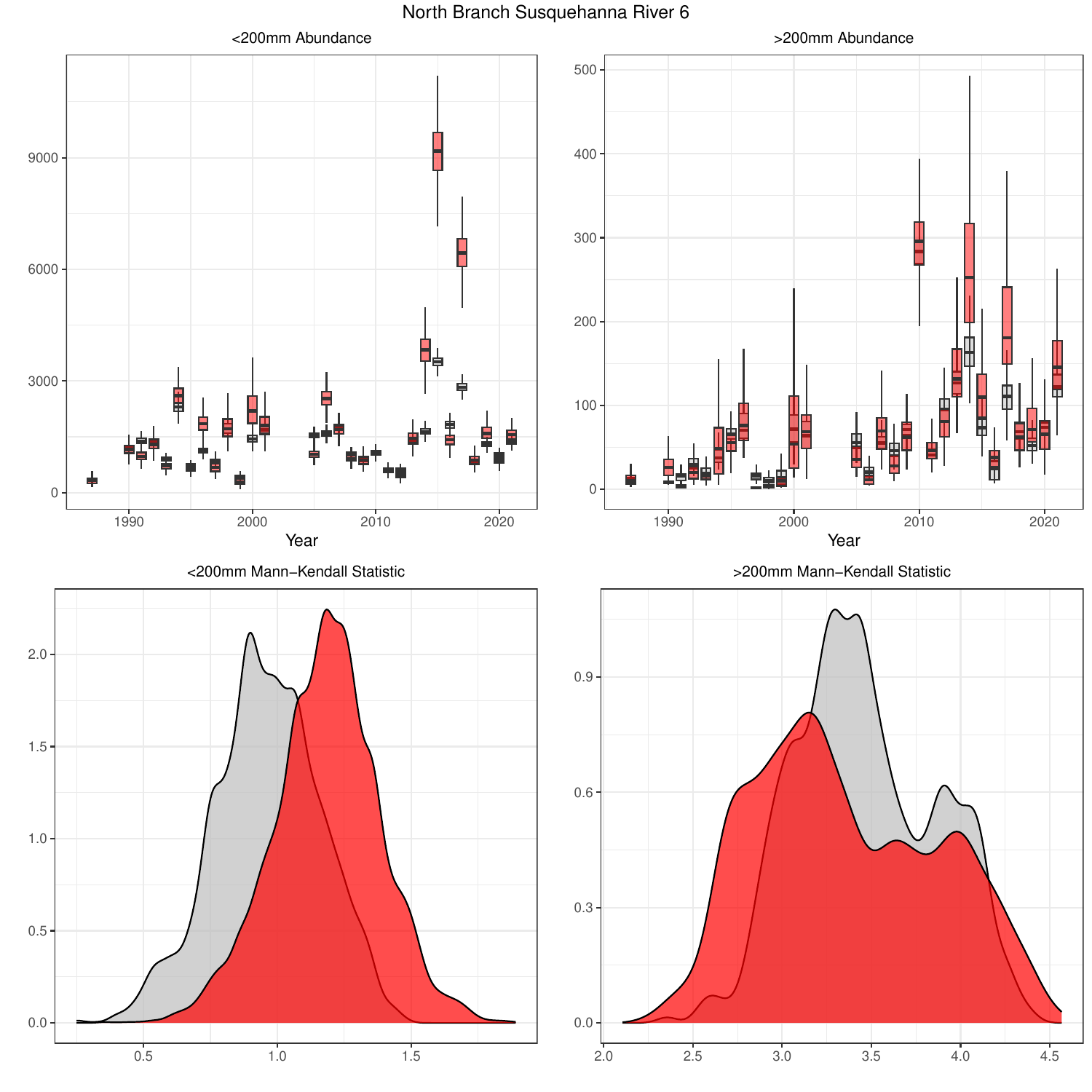}
    \caption{Model output for North Branch Susquehanna River Segment 6 CPUE data. The top two panels are posterior distributions for $\tilde N_{tj}$ (grey) and $\tilde N_{tj}^*$ (red). The bottom two panels are posterior distributions for $U(\tilde N_{tj})$ (grey) and $U(\tilde N_{tj}^*)$ (red).}
    \label{fig:nbsusq6}
\end{figure}

\begin{figure}
    \centering
    \includegraphics[width=0.8\linewidth]{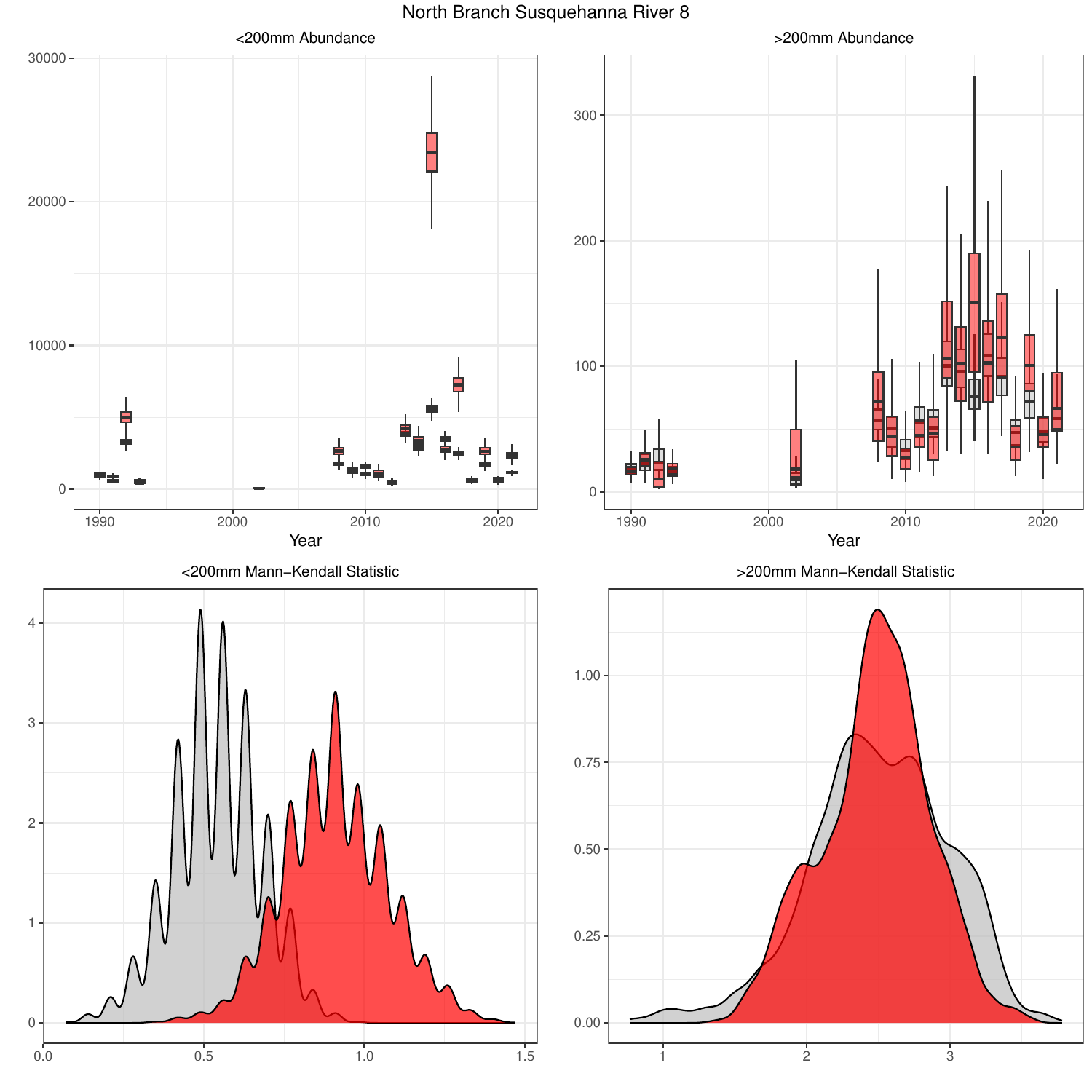}
    \caption{Model output for North Branch Susquehanna River Segment 8 CPUE data. The top two panels are posterior distributions for $\tilde N_{tj}$ (grey) and $\tilde N_{tj}^*$ (red). The bottom two panels are posterior distributions for $U(\tilde N_{tj})$ (grey) and $U(\tilde N_{tj}^*)$ (red).}
    \label{fig:nbsusq8}
\end{figure}

\begin{figure}
    \centering
    \includegraphics[width=0.8\linewidth]{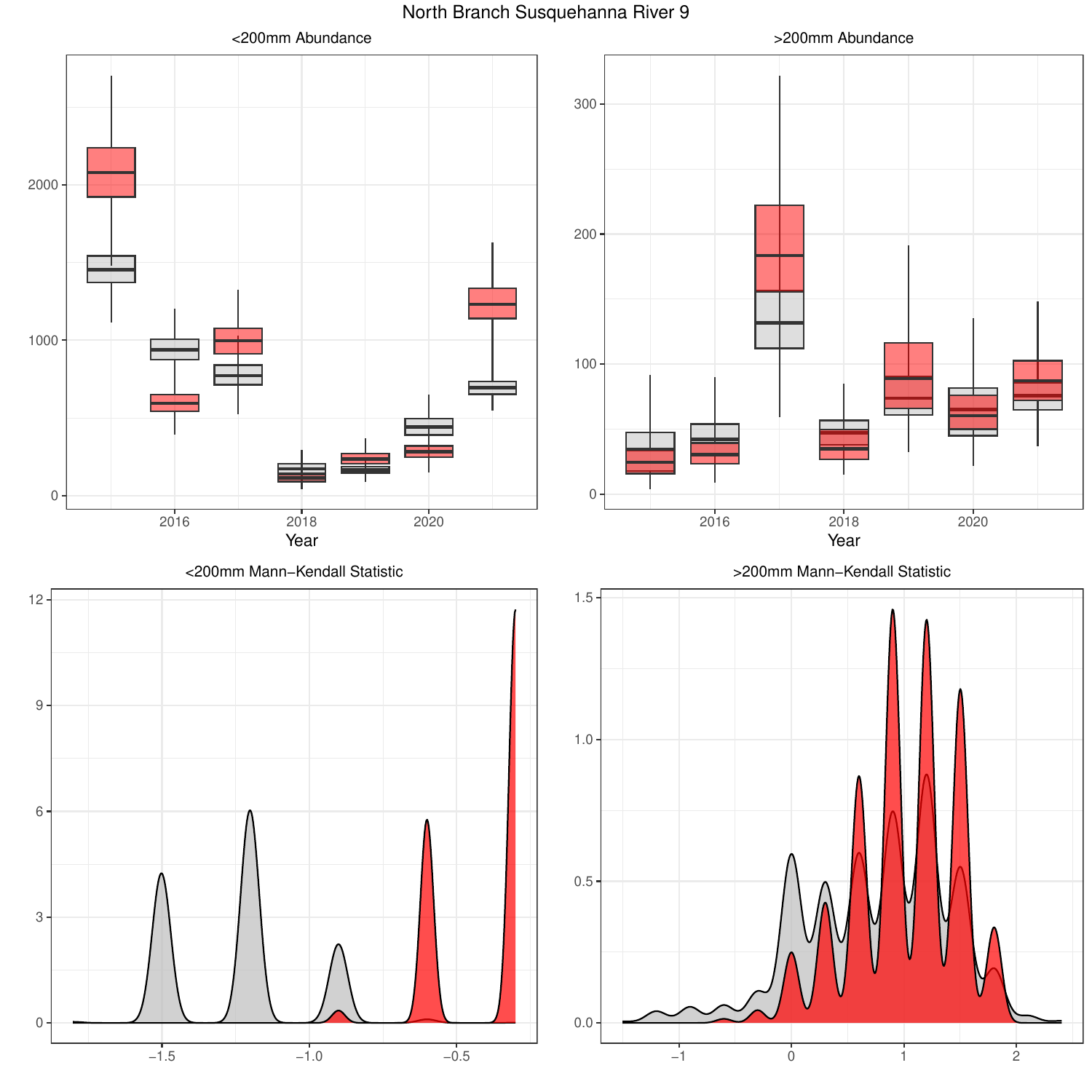}
    \caption{Model output for North Branch Susquehanna River Segment 9 CPUE data. The top two panels are posterior distributions for $\tilde N_{tj}$ (grey) and $\tilde N_{tj}^*$ (red). The bottom two panels are posterior distributions for $U(\tilde N_{tj})$ (grey) and $U(\tilde N_{tj}^*)$ (red).}
    \label{fig:nbsusq9}
\end{figure}

\begin{figure}
    \centering
    \includegraphics[width=0.8\linewidth]{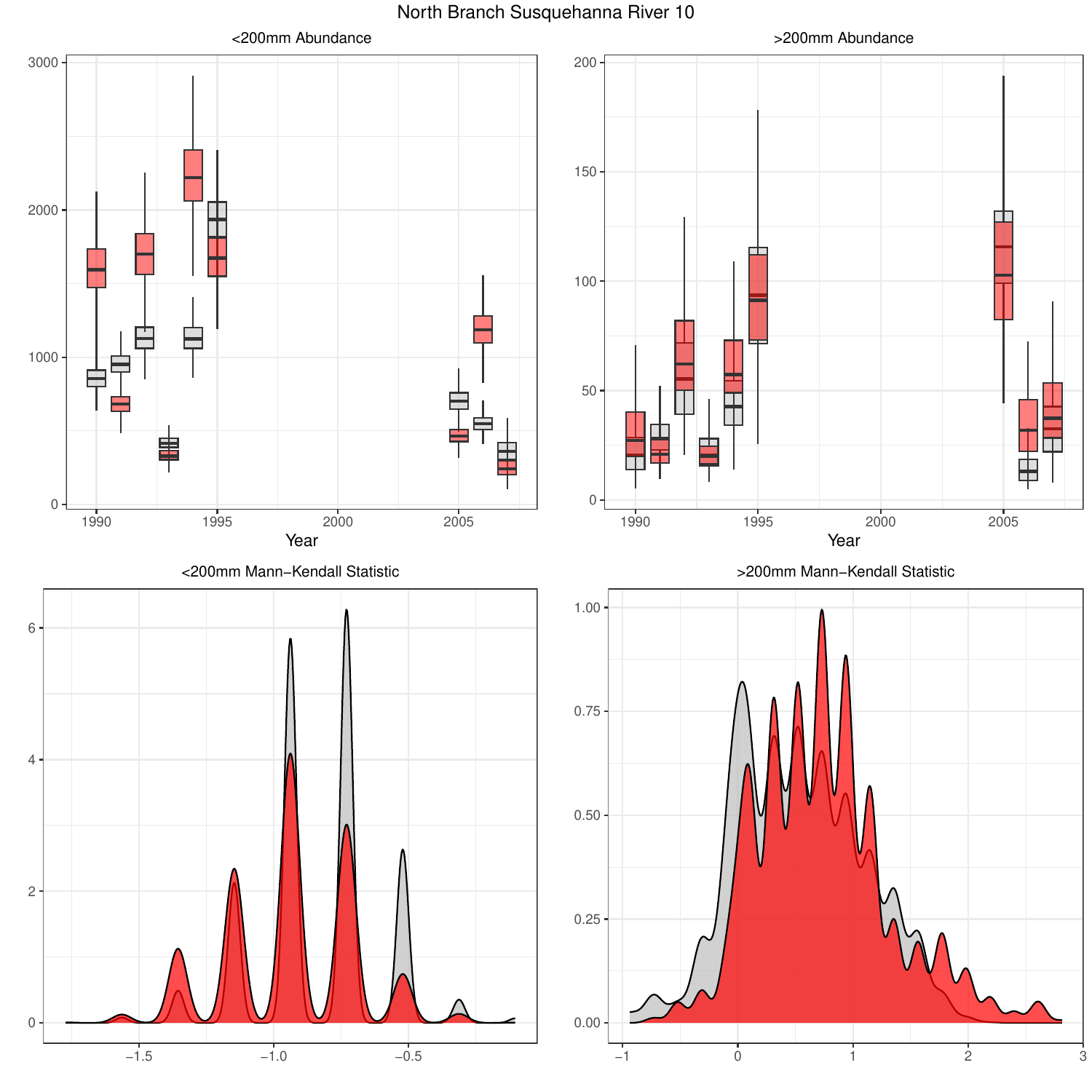}
    \caption{Model output for North Branch Susquehanna River Segment 10 CPUE data. The top two panels are posterior distributions for $\tilde N_{tj}$ (grey) and $\tilde N_{tj}^*$ (red). The bottom two panels are posterior distributions for $U(\tilde N_{tj})$ (grey) and $U(\tilde N_{tj}^*)$ (red).}
    \label{fig:nbsusq10}
\end{figure}

\begin{figure}
    \centering
    \includegraphics[width=0.8\linewidth]{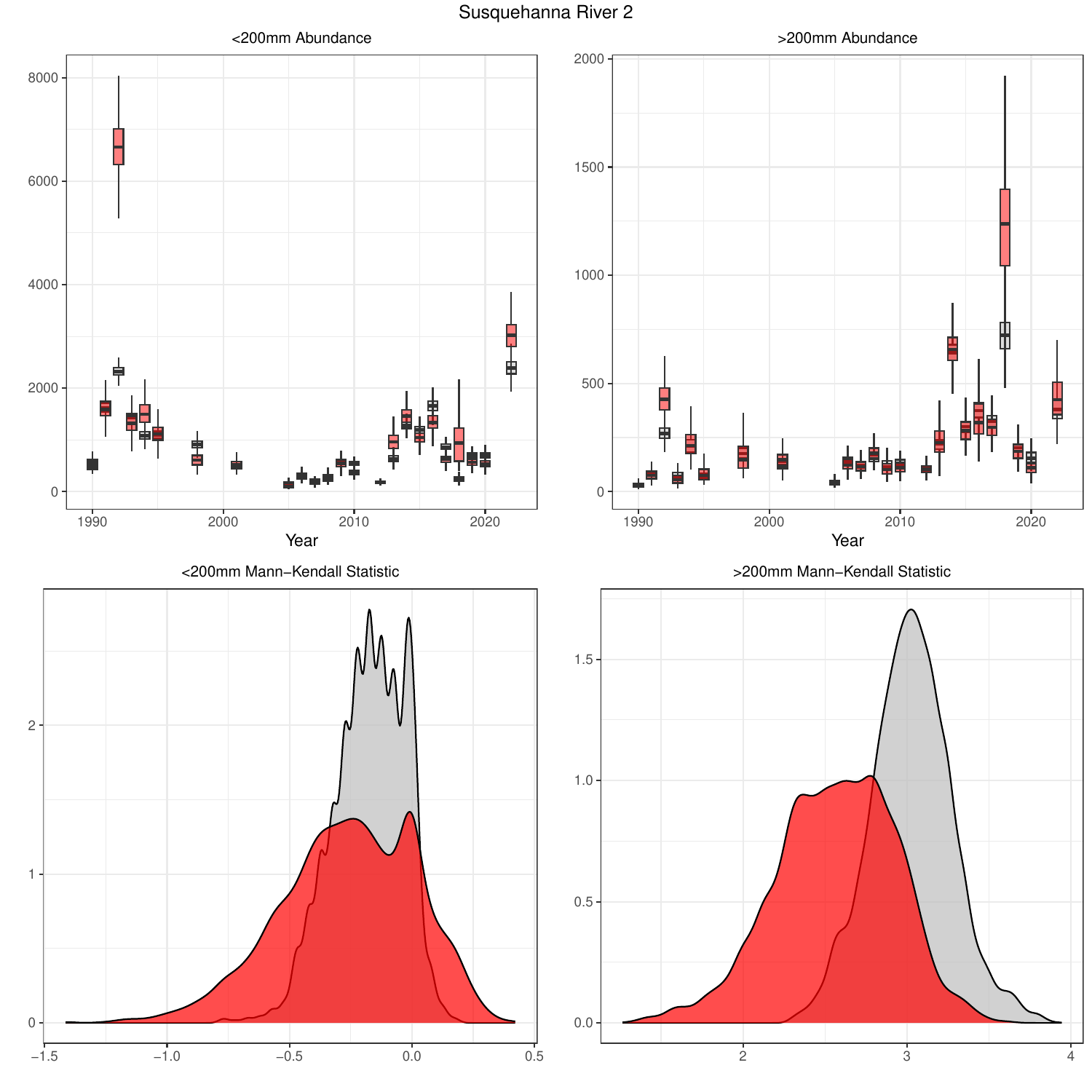}
    \caption{Model output for Susquehanna River Segment 2 CPUE data. The top two panels are posterior distributions for $\tilde N_{tj}$ (grey) and $\tilde N_{tj}^*$ (red). The bottom two panels are posterior distributions for $U(\tilde N_{tj})$ (grey) and $U(\tilde N_{tj}^*)$ (red).}
    \label{fig:susq2}
\end{figure}

\begin{figure}
    \centering
    \includegraphics[width=0.8\linewidth]{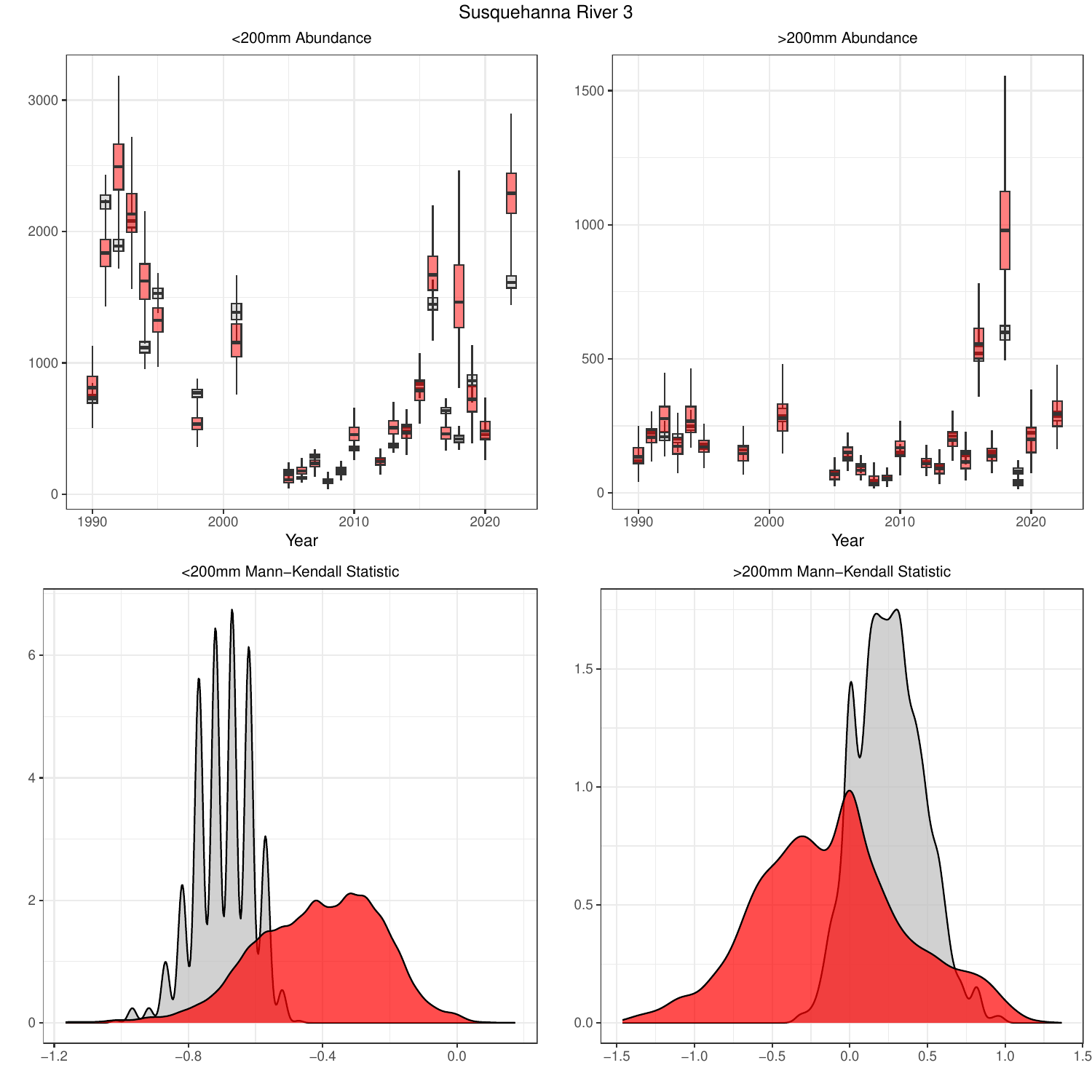}
    \caption{Model output for Susquehanna River Segment 3 CPUE data. The top two panels are posterior distributions for $\tilde N_{tj}$ (grey) and $\tilde N_{tj}^*$ (red). The bottom two panels are posterior distributions for $U(\tilde N_{tj})$ (grey) and $U(\tilde N_{tj}^*)$ (red).}
    \label{fig:susq3}
\end{figure}

\begin{figure}
    \centering
    \includegraphics[width=0.8\linewidth]{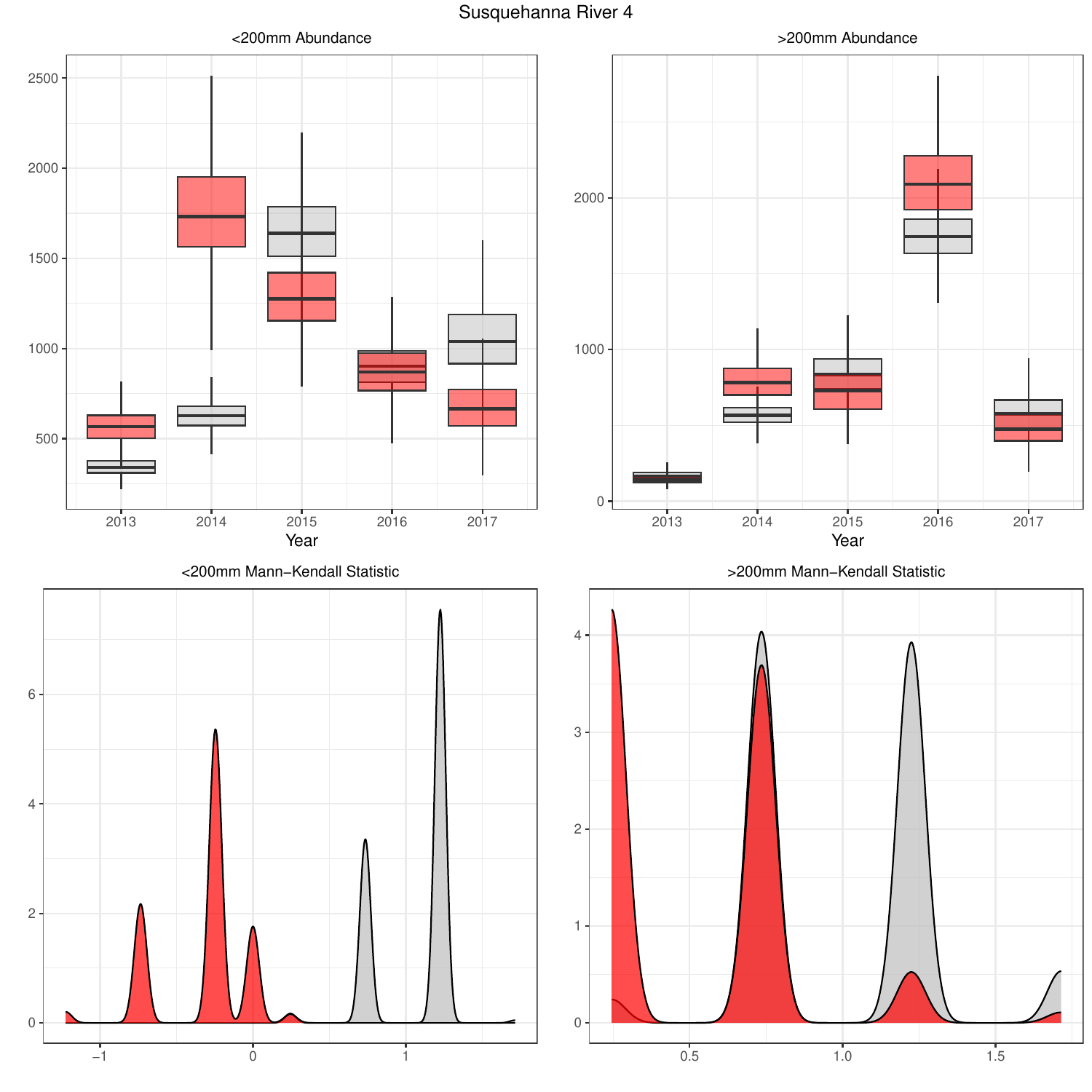}
    \caption{Model output for Susquehanna River Segment 4 CPUE data. The top two panels are posterior distributions for $\tilde N_{tj}$ (grey) and $\tilde N_{tj}^*$ (red). The bottom two panels are posterior distributions for $U(\tilde N_{tj})$ (grey) and $U(\tilde N_{tj}^*)$ (red).}
    \label{fig:susq4}
\end{figure}

\begin{figure}
    \centering
    \includegraphics[width=0.8\linewidth]{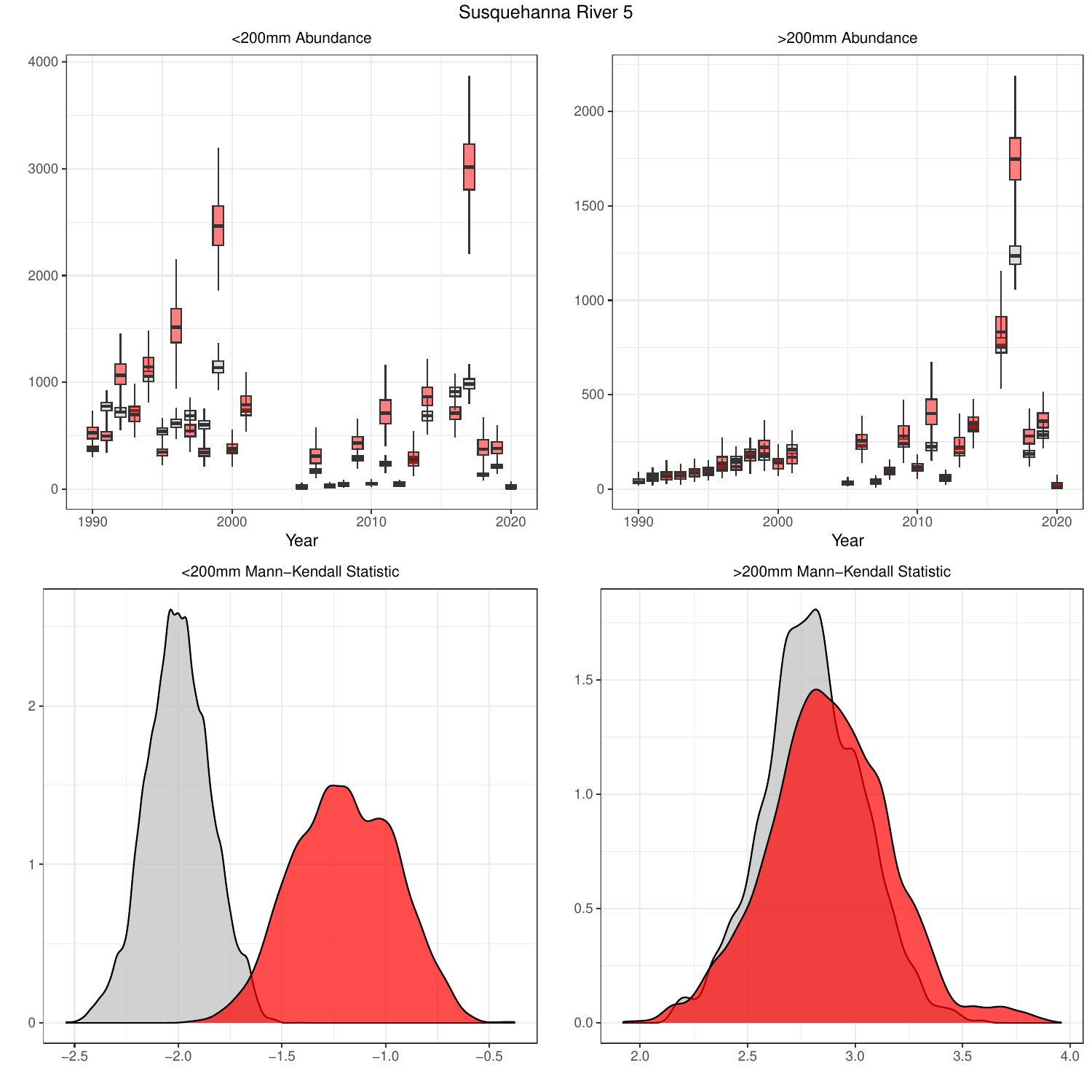}
    \caption{Model output for Susquehanna River Segment 5 CPUE data. The top two panels are posterior distributions for $\tilde N_{tj}$ (grey) and $\tilde N_{tj}^*$ (red). The bottom two panels are posterior distributions for $U(\tilde N_{tj})$ (grey) and $U(\tilde N_{tj}^*)$ (red).}
    \label{fig:susq5}
\end{figure}

\begin{figure}
    \centering
    \includegraphics[width=0.8\linewidth]{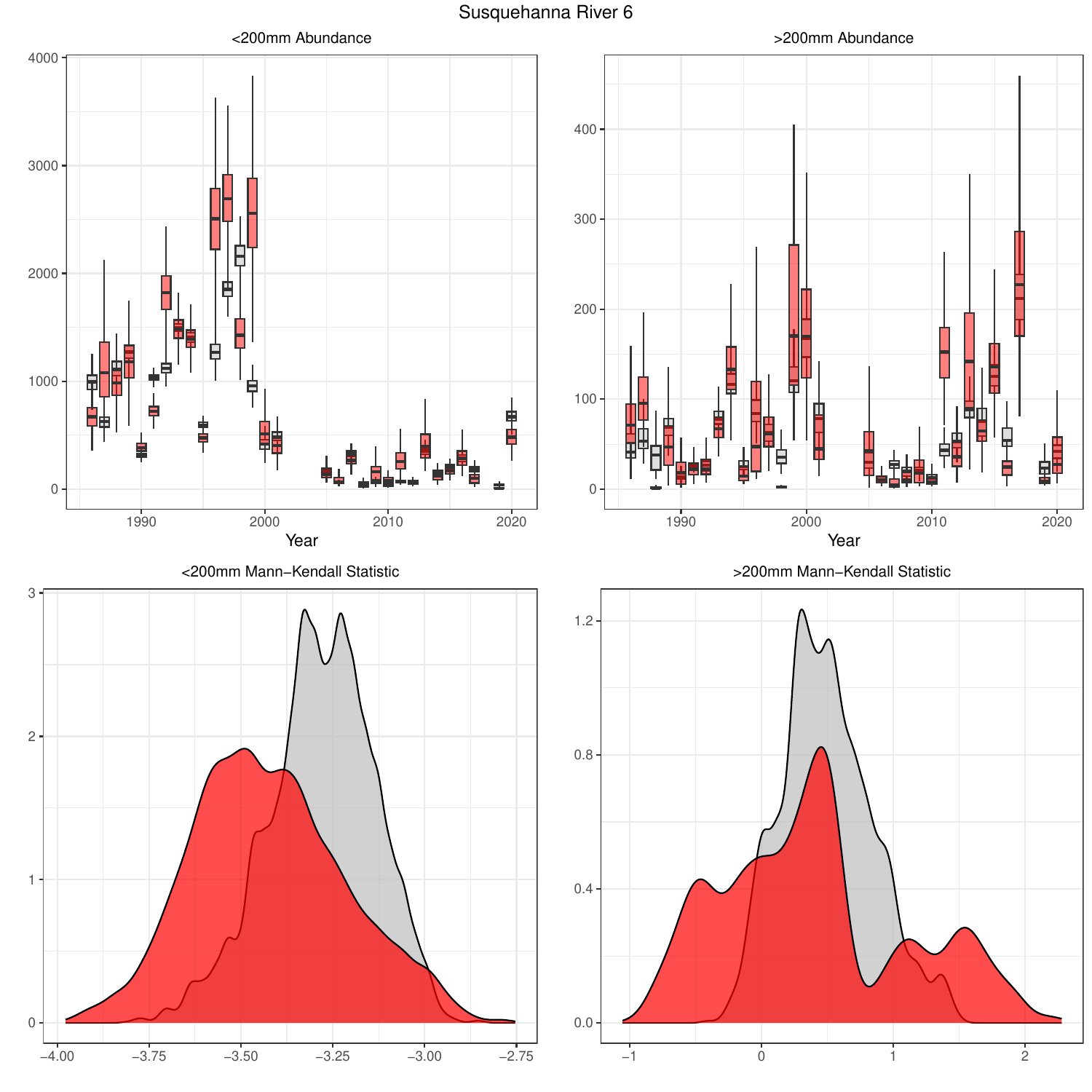}
    \caption{Model output for Susquehanna River Segment 6 CPUE data. The top two panels are posterior distributions for $\tilde N_{tj}$ (grey) and $\tilde N_{tj}^*$ (red). The bottom two panels are posterior distributions for $U(\tilde N_{tj})$ (grey) and $U(\tilde N_{tj}^*)$ (red).}
    \label{fig:susq6}
\end{figure}

\begin{figure}
    \centering
    \includegraphics[width=0.8\linewidth]{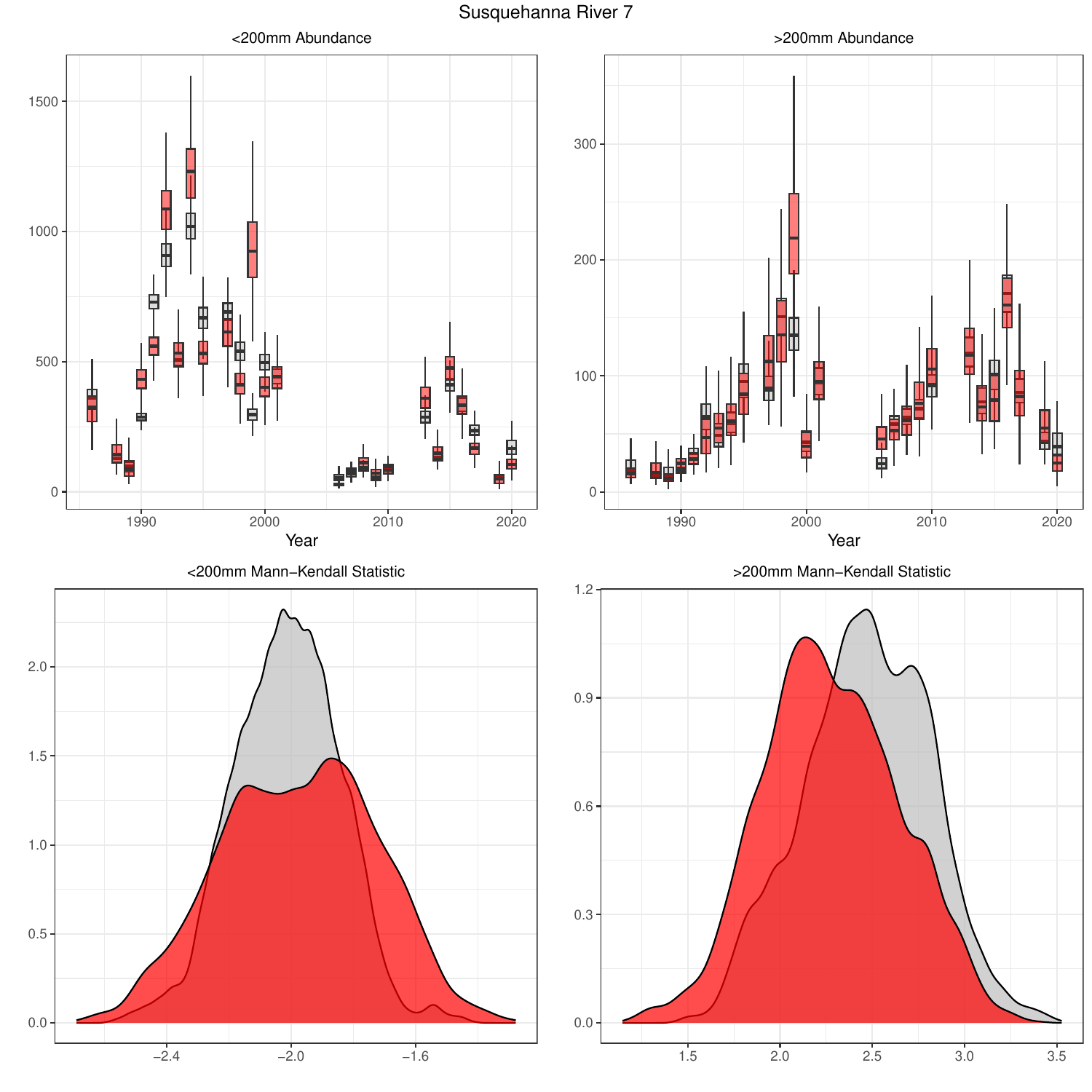}
    \caption{Model output for Susquehanna River Segment 7 CPUE data. The top two panels are posterior distributions for $\tilde N_{tj}$ (grey) and $\tilde N_{tj}^*$ (red). The bottom two panels are posterior distributions for $U(\tilde N_{tj})$ (grey) and $U(\tilde N_{tj}^*)$ (red).}
    \label{fig:susq7}
\end{figure}

\begin{figure}
    \centering
    \includegraphics[width=0.8\linewidth]{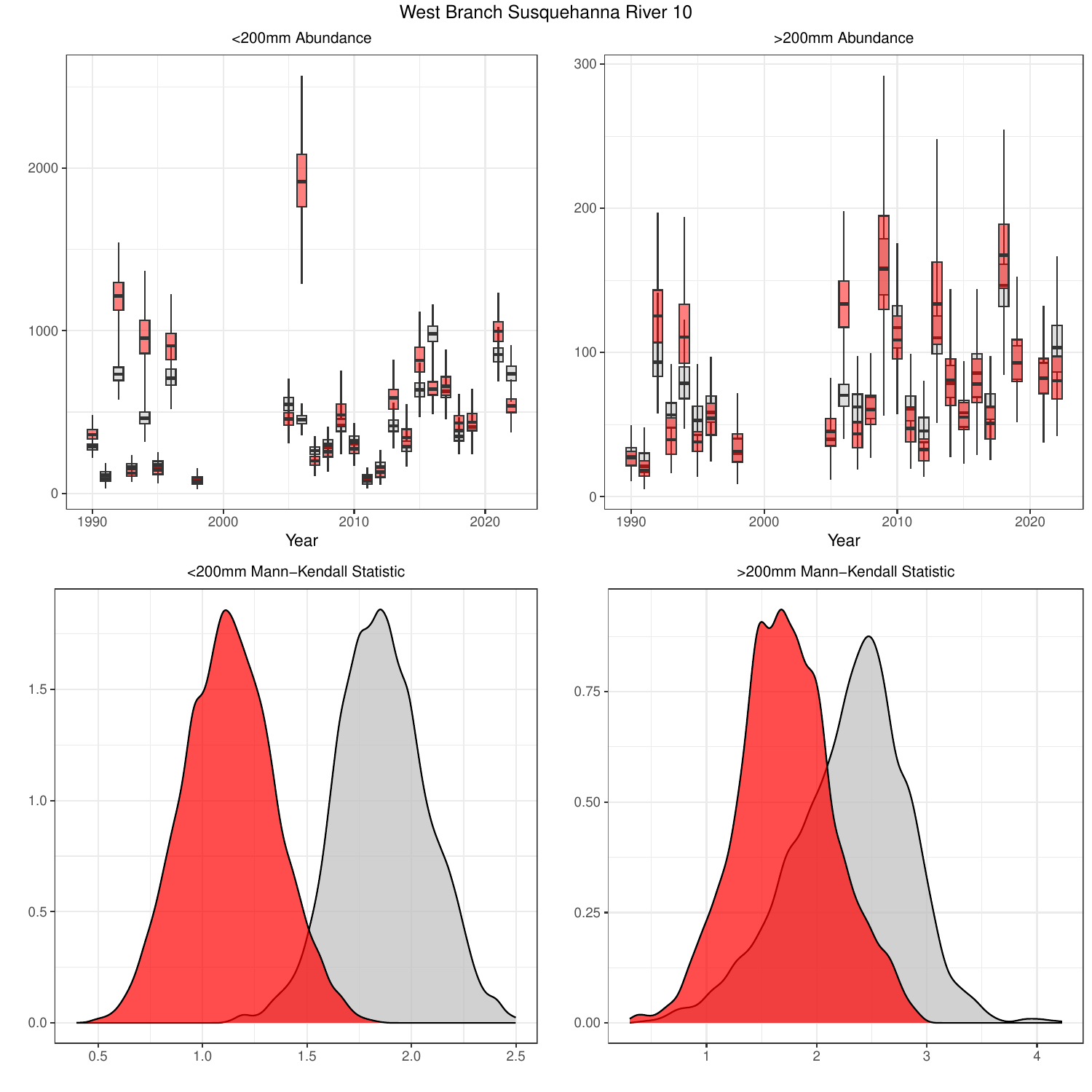}
    \caption{Model output for West Branch Susquehanna River Segment 10 CPUE data. The top two panels are posterior distributions for $\tilde N_{tj}$ (grey) and $\tilde N_{tj}^*$ (red). The bottom two panels are posterior distributions for $U(\tilde N_{tj})$ (grey) and $U(\tilde N_{tj}^*)$ (red).}
    \label{fig:wbsusq10}
\end{figure}

\label{lastpage}

\end{document}